\documentclass[a4paper,onecolumn,superscriptaddress,11pt,accepted=2018-08-15]{quantumarticle}
\pdfoutput=1
\usepackage[utf8]{inputenc}
\usepackage[english]{babel}
\usepackage[T1]{fontenc}
\usepackage{amsmath,amssymb,amsfonts}
\usepackage{hyperref}

\usepackage{tikz}
\usepackage{lipsum}

\usepackage[square,comma,numbers,sort&compress]{natbib}

\newcommand{\lambdabar}{{\mkern0.75mu\mathchar '26\mkern -9.75mu\lambda}}

\definecolor{mygrey}{gray}{0.35}
\definecolor{myblue}{rgb}{0.2,0.2,0.8}
\definecolor{myzard}{cmyk}{0,0,0.05,0}
\definecolor{mywhite}{rgb}{1,1,1}
\definecolor{mywhite}{rgb}{1,1,1}
\definecolor{myred}{rgb}{1,0.,0.3}

\newcommand{\bra}[1]{\left\langle #1\right|}
\newcommand{\ket}[1]{\left| #1\right\rangle}

\newcommand{\mean}[1]{\langle #1\rangle}

\newcommand\cc{\mathbf{c}}
\newcommand\dd{\mathbf{d}}

\newcommand\nn{\mathbf{n}}
\newcommand\mm{\mathbf{m}}

\newcommand\kk{\mathbf{k}}
\newcommand\qq{\mathbf{q}}
\newcommand\pp{\mathbf{p}}

\newcommand\bath{\mathrm{bath}}

\newcommand\EE{\mathbf{E}}

\newcommand\rr{\mathbf{r}}
\newcommand\RR{\mathbf{R}}

\newcommand\PP{\mathbb{P}}
\newcommand\QQ{\mathbb{Q}}

\newcommand\KK{\mathrm{K}}
\newcommand\hc{\mathrm{H.c.}}

\newcommand\intt{\mathrm{int}}

\newcommand\CS{\mathrm{CS}}
\newcommand\FCC{\mathrm{FCC}}
\newcommand\BCC{\mathrm{BCC}}

\newcommand\BCD{\mathrm{BCD}}
\newcommand\BC{\mathrm{BC}}
\newcommand\BS{\mathrm{BS}}
\newcommand\UP{\mathrm{UP}}

\begin{document}

\title{Non-Markovian Quantum Optics with Three-Dimensional State-Dependent Optical Lattices }
 \author{A. Gonz\'{a}lez-Tudela}
 \email{a.gonzalez.tudela@csic.es}
 \affiliation{Max-Planck-Institut f\"{u}r Quantenoptik Hans-Kopfermann-Str. 1. 85748 Garching, Germany }
 \affiliation{Instituto de F\'isica Fundamental IFF-CSIC, Calle Serrano 113b, Madrid 28006, Spain.}
 \author{J. I. Cirac}
 \affiliation{Max-Planck-Institut f\"{u}r Quantenoptik Hans-Kopfermann-Str. 1. 85748 Garching, Germany }

\begin{abstract}
 Quantum emitters coupled to structured photonic reservoirs experience unconventional individual and collective dynamics emerging from the interplay between dimensionality and non-trivial photon energy dispersions.  In this work, we systematically study several paradigmatic three dimensional structured baths with qualitative differences in their bath spectral density. We discover non-Markovian individual and collective effects absent in simplified descriptions, such as perfect subradiant states or long-range anisotropic interactions. Furthermore, we show how to implement these models using only cold atoms in state-dependent optical lattices and how this unconventional dynamics can be observed with these systems.
\end{abstract}

\maketitle

\section{Introduction \label{sec:intro}}

Initially motivated by overcoming the figures of merit of standard quantum optical setups, there exists a growing interest in integrating quantum emitters with nanophotonic structures ~\cite{vetsch10a,thompson13a,goban13a,beguin14a,lodahl15a,sipahigil16a,corzo16a,sorensen16a,solano17a}. The confined photons in these structures display highly structured energy dispersions, whose interplay with the dimensionality induces qualitatively new phenomena in the individual and collective quantum emitter (QE) dynamics. To name a few examples, these structured photonic reservoirs give rise to the emergence of atom-photon bound states~\cite{bykov75a,john90a,kurizki90a,tanaka06a,calajo16a,shi16a}, novel super/subradiant phenomena~\cite{facchi16a,galve17a,gonzaleztudela17a,asenjogarcia17a,shahmoon17a,glaetzle17a,perczel17a,albrecht18a} or coherent non-local interactions~\cite{douglas15a,gonzaleztudela15c,shahmoon16a}, among others. On top of that, these structured baths also display non-analytical spectral regions 
in their density 
of states, e.g., band edges, where 
non-trivial dynamics emerge~\cite{john94a,tong10a,longo10a,garmon13a,redchenko14a,lombardo14a,sanchezburillo17a} beyond what could be observed in unstructured baths~\cite{lehmberg70a}. This phenomena is typically referred to as non-Markovian dynamics, since standard perturbative treatments like Born-Markov master equations~\cite{gardiner_book00a} fail. The failure of conventional methods poses a great theoretical challenge for the characterization of these systems, which together with the possibility of observing exotic phenomena, makes the study of such non-Markovian dynamics still nowadays a very active research area~(see Ref~\cite{devega17a} and references therein for a recent review on the subject).
6
The exciting prospects of finding qualitative different phenomena compared to standard quantum optical systems have also attracted the attention of other communities beyond the optical regime, like circuit QED~\cite{liu17a,sundaresan18a,mirhosseini18a} or atoms in state-dependent optical lattices~\cite{devega08a,navarretebenlloch11a,krinner18a,stewart17a}, allowing one to enter in a completely new parameter regime as compared to the optical implementations. As an illustration of the potential of these new platforms, they have already succeeded observing single-photon bound-state physics associated to 1D band-edges in frequency~\cite{liu17a} and time domain~\cite{krinner18a}, a long-standing prediction in the optical regime~\cite{bykov75a,john90a,kurizki90a,john94a}, which took more than 40 years to be observed~\cite{hood16a}.

So far, most studies of these unconventional setups are devoted to one and two-dimensional baths, while the physics emerging from three-dimensional (3D) structured baths remains largely unexplored due to the theoretical and experimental challenges to study them. From the theoretical side, 3D structured baths are challenging due to the large Hilbert space to be considered. This is why most theoretical studies of 3D baths use simplified models, e.g., isotropic dispersions~\cite{john90a,kurizki90a,devega08a,navarretebenlloch11a}. From the experimental point of view, they also represent an outstanding challenge, since the integration and addressing of QEs embedded in three dimensional systems will be extremely difficult in nanophotonics or circuit QED implementations. For all those reasons, the characterization and implementation of these unconventional 3D structured baths represent one of the current frontiers of quantum 
optical studies.

In this manuscript, we both i) analyze the individual and collective dynamics of QEs coupled to several paradigmatic 3D baths, discovering non-Markovian phenomena qualitatively different from other reservoirs; and ii) we propose a setup based on state-dependent optical lattices where to observe them~\cite{devega08a,navarretebenlloch11a}. In particular, we consider several nearest-neighbour tight-binding baths with different geometries: cubic simple (CS), body-centered cubic (BCC), face-centered cubic (FCC) and diamond lattice~\cite{ashcroft76a}. All these baths show distinctive features in their density of states, which translates in very different individual and collective QE dynamics. We discover a plethora of effects absent in simplified descriptions~\cite{john90a,kurizki90a,devega08a,navarretebenlloch11a}, such as:
\begin{itemize}
 \item Long-time reversible dynamics for a single QE spectrally tuned at the middle of the band for CS geometries;
 \item Directional emission and perfect subradiance with QEs in BCC geometries;
 \item Robust 3D bound states leading to anisotropic coherent interactions between QEs in FCC lattices;
 \item Purely long-range coherent interactions between QEs in diamond geometries;
\end{itemize}
among many other effects. We use both numerical and analytical techniques, which allows us to extract the scalings with the system parameters of these features analytically. Regarding the implementation with state-dependent optical lattices, i) we provide the laser configurations required to implement these structured baths with optical dipole traps, ii) we analyze the estimated timescales of the dynamics, and finally iii) we study the deviations from the idealized nearest neighbour description of the bath dynamics, focusing on their impact in the bath density of states. Overall, we show how the observation of the discovered non-Markovian 3D phenomena can potentially be observed with three-dimensional state-dependent optical lattices.

The outline of this manuscript reads as follows: in Section~\ref{sec:setup} we discuss the general features of the setup that we study along this manuscript. Here, we also explain how to implement these models using cold atoms in state-dependent optical lattices, since it is currently the more realistic platform where to implement them. Then, to make the manuscript self-contained, in Section~\ref{sec:theory} we briefly review the theoretical techniques we use to study the problem. In Section~\ref{sec:bath}, we give an overview of the properties of the different baths considered  along the manuscript, explaining both the lattice geometry and the expected density of states in the thermodynamic limit. Then, in Sections~\ref{sec:CS}-\ref{sec:diamond} we calculate the quantum emitter dynamics for the aforementioned baths of interest, focusing on non-Markovian phenomena and emphasizing the differences with respect to other types of structured reservoirs. After presenting the emergent phenomena, in Section~\ref{sec:OLimplementation} we go back to the implementation discussion and show the laser configuration required to implement the bath Hamiltonians studied along this 
manuscript in optical lattices. Furthermore, we calculate the band structure of the proposed lattices to estimate how well they describe the toy models considered along the manuscript, which we characterize by looking at the density of states of the obtained model. Finally, in Section~\ref{sec:conclusion} we summarize the main results of the manuscript, and point to future work directions.

\section{General quantum optical setup and its implementation with state-dependent optical lattices\label{sec:setup}}

In this Section we describe the general model that we consider along this manuscript and its natural implementation with cold atoms in state dependent optical lattices. 

\subsection{General model}

We are interested in the dynamics of $N_e$ QEs, which we describe as two-level systems, $\{\ket{g}_j,\ket{e}_j\}_{j=1}^{N_e}$, whose intrinsic dynamics is given by the following Hamiltonian (we use $\hbar\equiv 1$ along the manuscript):
\begin{equation}
\label{eq:Hsingle}
 H_S=\omega_e \sum_{j=1}^{N_e} \sigma_{ee}^j\,,
\end{equation}
where we use the notation $\sigma_{\alpha\beta}^j=\ket{\alpha}_j\bra{\beta}_j$ for the spin operator of the $j$-th QE, and $\omega_e$ is the transition frequency of the QE optical transition that couples to the bath modes. 

For the bath description, we take a discretized version with a finite number of bosonic modes. We consider baths of linear size $N$, such that the total number of sites is $N\times N\times N$. For the baths that we consider along this manuscript, one has to distinguish two situations, namely, the single and two-band configurations. The single-band model corresponds to a configuration where the bath geometry can be described as a simple Bravais lattice~\cite{ashcroft76a}. This means that the position of each bosonic mode, with creation/annihilation operators $a^\dagger_\nn/a_\nn$, is given by three integer numbers: $\nn=(n_1,n_2,n_3)=\sum_{i=1}^3 n_i \cc_i$, with $n_i\in (0,\dots,N-1)$, which describes the displacement in terms of the primitive vectors, $\cc_i$, expanding the lattice in real space. Unless stated otherwise, we assume that each bosonic mode only couples to its nearest-neighbour, with strength $J$, that we will take as the unit of energy. The number and positions of the nearest neighbours 
depend on the particular bath geometry, as we will show in Section~\ref{sec:bath}. Using all these assumptions, the bath Hamiltonian can be generally written as follows:
\begin{equation}
\label{eq:Hbath}
 H_\bath=\omega_a \sum_\nn a^\dagger _\nn a_\nn-J \sum_{\mean{\nn,\mm}}a^\dagger_\mm a_\nn=\sum_\kk \omega(\kk) \hat{a}^\dagger_\kk \hat{a}_\kk\,,
\end{equation}
where $\omega_a$ is the bosonic mode frequency which we assume to be equal for all the bath sites. In the last step we have assumed periodic boundary conditions for the bath modes and introduced the operators $\hat{a}_\kk=\frac{1}{N^{3/2}}\sum_\nn e^{i \kk\cdot \nn} a_\nn$, which diagonalize the bath Hamiltonian in momentum space. Notice, we are using the hat notation, $\hat{\cdot}$, to distinguish the bath operators in position/momentum space. The momenta $\kk$ are also described by three numbers $\kk=(k_1,k_2,k_3)=\sum_i k_i \dd_i$, being $\dd_i$ are the primitive vectors of the reciprocal lattice, which satisfy: $\cc_i\cdot \dd_j=\delta_{ij}$. As we show in Section~\ref{sec:bath}, this single-band situation describes the cases of the CS, BCC and FCC lattices. 

There are bath geometries, however, which cannot be described as simple Bravais lattices~\cite{ashcroft76a}. For example, the diamond lattice is described by two interspersed FCC lattices (we explain this in detail in Section~\ref{sec:bath}). In those situations, one needs to upgrade the single-band model to a more general case which allows one to capture the exact dynamics of the bath. For this manuscript, we consider the simplest upgrade, namely, a two-band model consisting of two discretized $N\times N\times N$ lattices (denoted as A/B sublattices), with associated bosonic modes, $a_\nn,b_\nn$, with the same energy $\omega_a$, interacting through nearest neighbour coupling as follows:
\begin{widetext}
\begin{align}
\label{eq:Hbathtwo}
 H_{\bath,t}&=\omega_a \sum_\nn a^\dagger _\nn a_\nn+\omega_a\sum_\nn b^\dagger_\nn b_\nn-J \sum_{\mean{\nn,\mm}}\left(a^\dagger_\mm b_\nn+\hc\right)=\sum_\kk \left(\omega_{u}(\kk) \hat{u}^\dagger_\kk \hat{u}_\kk+\omega_{l}(\kk)\hat{l}^\dagger_\kk \hat{l}_\kk\right)\,.
\end{align}
\end{widetext}

Here $\hat{u}_\kk, \hat{l}_\kk$ are the operators that diagonalize the bath Hamiltonian in momentum space, with eigenenergies: $\omega_{u,l}(\kk)$ respectively. This is obviously not the most general two-band configuration, since it may also occur that there exist hoppings between the AA/BB sites, or there is some energy off-set between sublattices. However, since the idealized diamond lattice is described by a Hamiltonian like in Eq.~\ref{eq:Hbathtwo}, we restrict ourselves to this situation in this manuscript. In both the single and two band models, it is convenient to move to a rotating frame with $\omega_a$, where $H_B\rightarrow H_B-\omega_a\sum_\nn a^\dagger _\nn a_\nn (-\omega_a\sum_\nn b^\dagger _\nn b_\nn )$  and $H_S=\Delta \sum_j \sigma_{ee}^j$, with $\Delta=\omega_e-\omega_a$.

\begin{figure}
\centering
\includegraphics[width=0.3\textwidth]{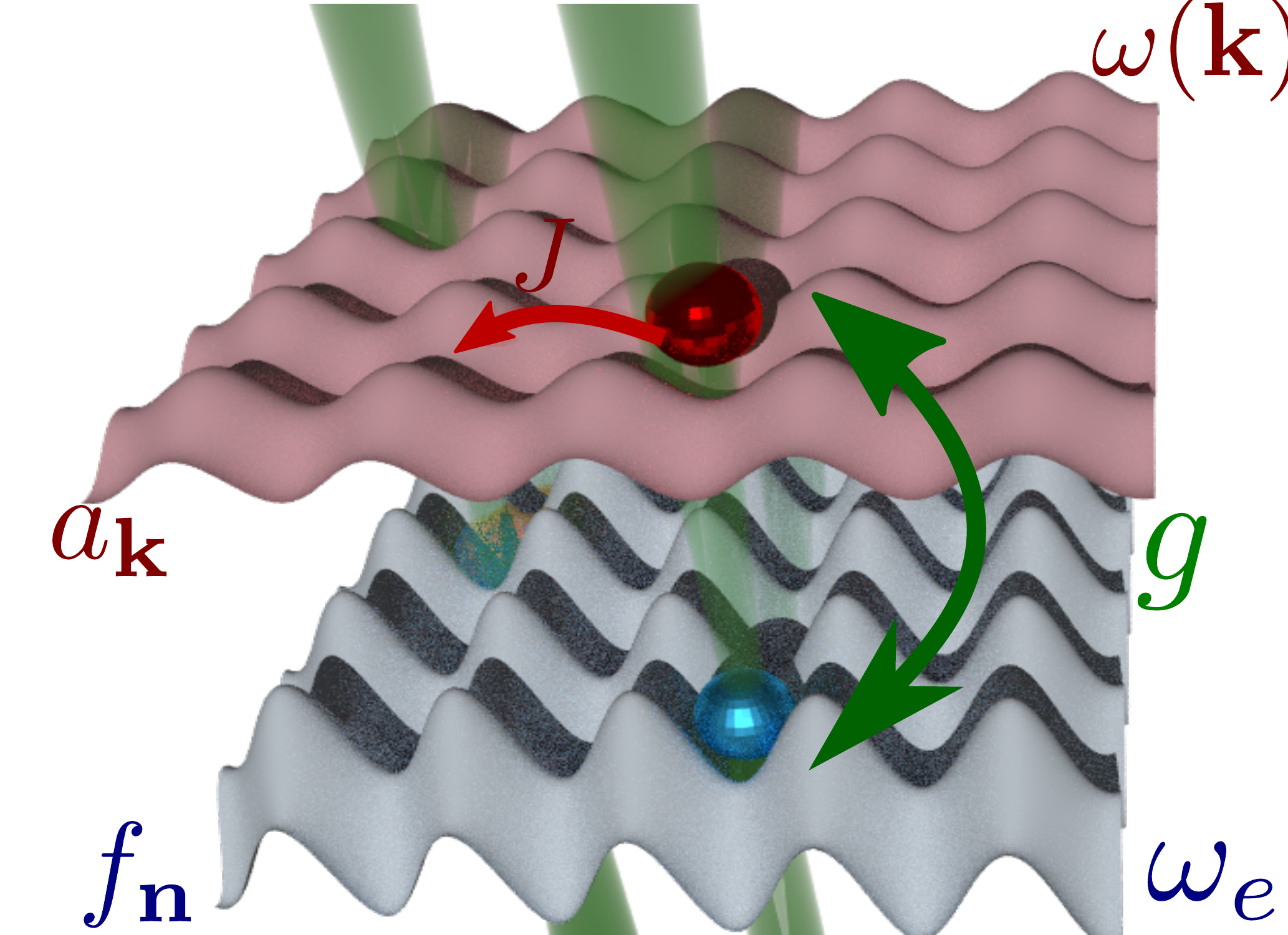}
\caption{Two dimensional view of the implementation of QED Hamiltonians with state-dependent optical lattices: two internal atomic states, labeled as $f$ and $a$, are trapped in very different trapping potentials. The $f/a$ atoms are trapped in a deep/shallow lattice playing the role of QEs/photon-like modes. They are connected through a two-photon Raman transition, or direct optical coupling for Alkaline-Earth atoms, at a rate $g$. }
\label{fig:1}
\end{figure}

The last ingredient is the system-bath interaction. For the sake of simplicity, we just assume a spatially local coupling between the QEs and the bath sites, since the spread of the QE wavefunction ultimately depends on the particular implementation considered. Mathematically, this local coupling assumption means that the $j$-th QE couples only to a bosonic mode at site $\nn_j$, such that the general light-matter interaction Hamiltonian generally reads:
\begin{align}
\label{eq:Hint}
 H_\intt&=g \sum_{j=1}^{N_e}\left(c_{\nn_j} \sigma_{eg}^j+\hc\right)=\frac{g}{N^{3/2}} \sum_\kk \sum_{j=1}^{N_e}\left(\hat{c}_{\kk} e^{i\kk\cdot\nn_j} \sigma_{eg}^j+\hc\right)\,.
\end{align}
where $c=a,b$ denotes a general bath operator, depending on whether the atom couples to an A/B site. Notice, that we have only written the excitation conserving terms in the light-matter Hamiltonian of Eq.~\ref{eq:Hint}. This is a safe assumption in the optical regime~\cite{cohenbook92a} and it will be an exact description in the cold-atoms implementation that we discuss in the next Section.

\subsection{General implementation with state-dependent optical lattices\label{sec:OL}}

In this subsection, we give a brief explanation on how to implement these structured 3D quantum optical models using cold atoms in state-dependent optical lattices. It will be mostly based on the proposal of Refs.~\cite{devega08a,navarretebenlloch11a}, which has been recently implemented with Rb atoms for 1D reservoirs~\cite{krinner18a}. 

The idea to mimic the quantum optical scenario with cold atoms is sketched in Fig.~\ref{fig:1}: we require two atomic internal states, that we label as $f$ and $a$, trapped in two very different potentials. One of them, e.g., the $f$ state, is trapped in a very deep potential such that its hopping to nearest neighbours is suppressed. We assume the trapping potential to be harmonic, with trapping frequency $\omega_e$, and restrict to the dynamics of the lowest motional state and in the collisional blockade regime where no more than one atom can be the trapped in each site. In this regime, the creation atomic operators $f^\dagger_\nn\approx \sigma_{eg}^\nn$ can be approximated by spin operators and therefore play the role of the QEs in our problem.

The other internal state, $a$, is trapped in a shallower potential, with trapping frequency $\omega_a$, such that they can hop to their nearest neighbours sites at a rate $J$. In the original proposal~\cite{devega08a,navarretebenlloch11a}, the atoms in this internal state were assumed to be free propagating particles with isotropic dispersion $\omega(\kk)\propto |\kk|^2$. Here instead, we consider that the $a$-modes feel the underlying potential, leading to a non-trivial energy dispersion $\omega(\kk)$ which depends on the geometry of the lattice considered.

Finally, the coupling between the two internal states can be obtained through two-photon Raman transitions~\cite{devega08a,navarretebenlloch11a,krinner18a}, or even through a direct optical transition in the case of Alkaline-Earth atoms~\cite{schreiber15a,daley08a,snigirev17a}, where there exist an optically excited state with lifetimes even longer than seconds. To mimic the dynamics of one or few excitations, we assume the laser addressing to be local, as depicted in Fig.~\ref{fig:1}. As shown in the original proposal~\cite{devega08a,navarretebenlloch11a}, these laser fields induce hopping between \emph{a/f} states, where the Hamiltonian reads:
\begin{align}
\label{eq:Hlas}
 H_\mathrm{las}= \sum_\kk \sum_{j=1}^{N_e} \left(g_\kk\hat{c}_{\kk} e^{i\kk\cdot\nn_j} \sigma_{eg}^j+\hc\right)\,,
\end{align}
where $g_\kk$ contains a $\kk$-dependence emerging from the finite size of the Wannier functions. This finite size of the wavefunction gives a natural cut-off in momentum space, whose main effect was shown to be a renormalization of the frequencies~\cite{devega08a,navarretebenlloch11a}. In the limit of very localized Wannier functions, and local laser addressing, $g_\kk\approx g/N^{3/2}$, it can be shown that $H_\mathrm{las}\approx H_\intt$. Remarkably, $H_\mathrm{las}$ conserves the total number of excitations of the system irrespective of $g$, unlike what happens in the optical regime. This allows one to explore parameter regimes very difficult to access with other platforms, including non-perturbative ones (such as $g>J$).

Finally, let us mention other advantages of the cold atom setup like: i) the possibility of single-site detection and addressing~\cite{sherson10a,weitenberg11a} to, e.g., observe in-situ the bath dynamics; ii) low decoherence rates ($<$Hz), compared to the Hamiltonian timescales $g,J\sim 10$ KHz; iii) importantly for this work, the possibility to simulate 3D models, something very difficult to implement with other platforms; and iv) the possibility to change $\omega(\kk)$ just by changing the laser configuration. For example, simulating CS baths can be done by simply sending 6 counter-propagating lasers in the X/Y/Z directions with orthogonal polarizations~\cite{bloch08a}, which results in a separable optical potential $V(\RR)\propto \cos^2(x)+\cos^2(y)+\cos^2(z)$. The configurations required for the rest of the lattices are less trivial. Thus, we leave the discussion on how to implement these optical lattices, and their resulting band structure, to Section~\ref{sec:OLimplementation}.

\begin{figure}
\centering
\includegraphics[width=0.9\textwidth]{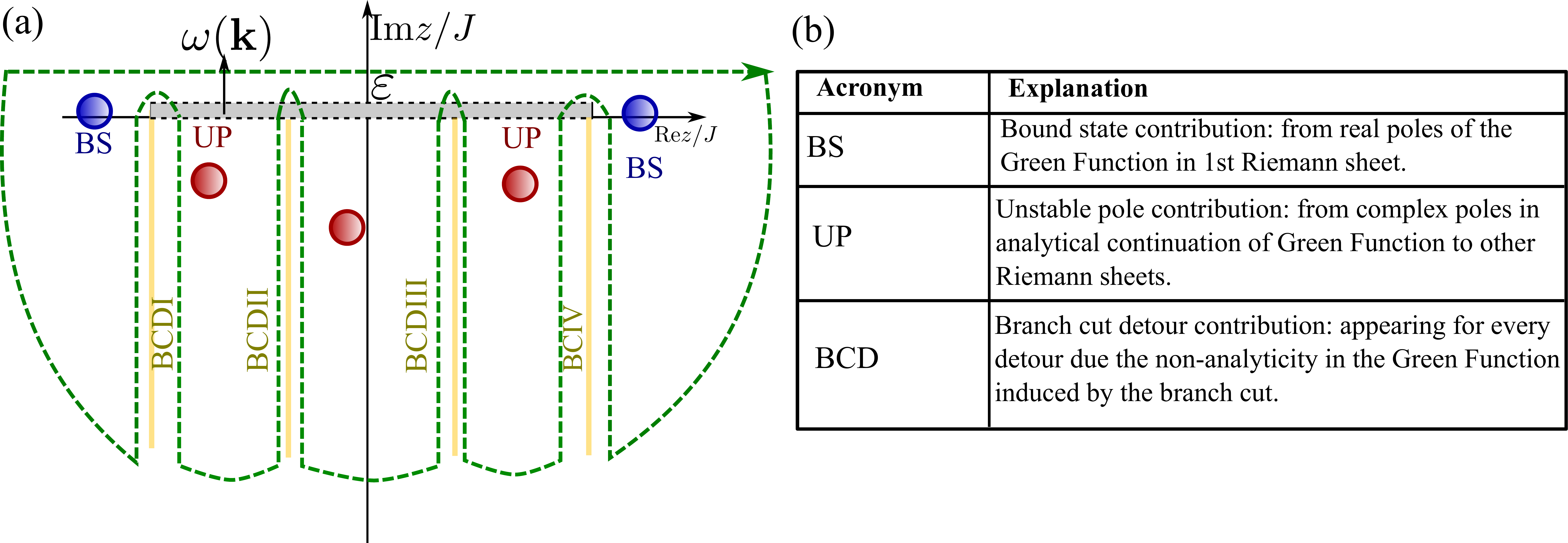}
\caption{(a) Example of a possible contour of integration to calculate $C_\alpha(t)$ as defined in Eq.~\ref{eq:inv}; One closes the contour with a semiarc in the lower complex plane. In the continuum limit, $G(z)$ develop non-analyticities and branch cuts that one must avoid with detours (in yellow) to be able to apply Residue Theorem. In this example there are four detours, although in general can be more or less (at least two for finite bands). Real (BS) and complex poles (UP), depicted in blue/red, will also contribute to the dynamics. (b) Summary of acronyms of the different contributions to the dynamics of $C_\alpha(t)$. }
\label{fig:2}
\end{figure}

\section{Theoretical tools\label{sec:theory}}

In this manuscript we mainly consider the spontaneous emission of a single excitation into the bath being emitted from one or few QEs as a first step to unravel the dynamics emerging from these unconventional 3D baths. This allows us to characterize both the individual and collective response in the linear regime. Importantly, the system-bath Hamiltonian, $H=H_S+H_\bath+H_\intt$, conserves the total number of excitations, such that we restrict to the single-excitation subspace where the calculations are more accessible. To calculate the dynamics we use two different and complementary approaches, which we describe in what follows.

\subsection{Analytical techniques}

The first one is an analytical approach based on the resolvent operator technique~\cite{cohenbook92a,nakazato96a}. This technique consists in calculating the Laplace transform of the time-evolution operator, $U(t)=e^{-i H t}$, which is the so-called Green-Function: $G(z)=\frac{1}{z-H}$. If $\{\ket{\alpha}\}$ is complete basis expanding $H$ and we are interested only in the dynamics of a particular subspace, e.g., $\PP=\ket{\beta}\bra{\beta}$,  one can show that the Green-Function in this subspace can be calculated as:
\begin{equation}
\label{eq:green}
 \PP G(z)\PP=\frac{1}{z-\PP H\PP -\Sigma(z)}\,,
\end{equation}
where $\PP$ is the projection in the subspace we are interested in, and $\Sigma(z)$ is the self-energy of the problem, that contains the effect of the interaction with the bath:
\begin{align}
 \label{eq:self}
\Sigma(z)=\PP H_\intt \frac{\QQ}{z-H_B} H_\intt \PP\,,
\end{align}
where $\QQ=1-\PP$. This method has two challenges: first, calculating the resolvent operator since for 3D baths is given, in the thermodynamic limit, by a 3D integral. For example, when only a single QE is coupled to a single-band bath it reads:
\begin{align}
 \label{eq:selfe}
 \Sigma_e(z)=\frac{g^2}{N^3}\sum_\kk \frac{1}{z-\omega(\kk)}=\frac{g^2}{(2\pi)^3}\iiint_\mathrm{BZ}\frac{d^3\kk}{z-\omega(\kk)}\,
\end{align}
where in the last equality we used: $\frac{1}{N^3}\sum_\kk\rightarrow \frac{1}{(2\pi)^3}\iiint d^3\kk$, which is the usual prescription to go to the thermodynamic limit. An important asset of this work is that we have analytical formulas for $\Sigma_e(z)$ for our baths of interests, which allows us to extract valuable information from the dynamics, such as scaling of decays, bound-state energies,\dots

The second challenge consists in moving from the Laplace to the time domain to get the dynamics. This can be done by calculating the inverse Laplace transform, which is done by solving an integral of the type:
\begin{align}
 \label{eq:inv}
 C_\alpha(t)=-\frac{1}{2\pi i}\int_{-\infty}^\infty dE G_\alpha(E+i0^+)e^{-i E t}\,,
\end{align}
where $C_\alpha(t)$ is the probability amplitude of a given state $\ket{\alpha}$. As sketched in Fig.~\ref{fig:2}(a), one possibility to solve these integrals is to find a closed a contour that contains the integral of Eq.~\ref{eq:inv} and apply complex integral techniques, such as Residue Theorem. However, depending on $\omega(\kk)$, $\Sigma(z)$, and consequently $G(z)$, may develop non-analytical regions in the thermodynamic limit, leading to branch cuts that have to be avoided with detours in the contour of integration (depicted in yellow in Fig.~\ref{fig:2}). Depending on $\omega(\kk)$ one must introduce as many detours as required to guarantee the analyticity of $G(z)$ in the whole closed region to be able to apply Residue Theory. Using this technique, one is able to decompose the dynamics of $C_\alpha(t)$ in different contributions (summarized in Table~\ref{fig:2}(b)):
\begin{itemize}

 \item Bound State (BS) contribution. They arise from real poles, $E_\BS\in \mathbb{R}$, of $G(z)$, whose energy lies out of the continuum. They give a contribution to the dynamics of the form: $R_\BS e^{-i E_\BS t}$, where $R_\BS$ is the residue of the pole. The choice of the BS notation is because the origin of these real poles are the emergence of photon bound states which localize around the QEs~\cite{john90a}.
 
 \item Unstable pole (UP) contribution. When taking the detours to avoid non-analytical regions in the domain of integration, one might need to analytically continue $G(z)$ to other Riemann sheets~\cite{nakazato96a}. These analytical continuations may show complex poles, $z_\UP$, with $\mathrm{Re}z_\UP\in \omega(\kk)$, and $\mathrm{Im}z_\UP\neq 0$ which also contribute to the dynamics as: $R_\UP e^{-i z_\UP t}$, being $R_\UP$ the residue associated to these poles.
 
\item Branch cut detour (BCD) contribution. As we explained, the branch cuts of $G(z)$ will force us to take detours at certain frequencies $E_\BC$. Typically, only band-edge detours are considered in the literature. However, we will see along this manuscript that they can also emerge in the middle of the band~\cite{gonzaleztudela17a,gonzaleztudela17b}. Since at both sides of the detour one must generally use a different analytical continuation of $G_e(z)$, this contribution can be generally written as:
 %
 \begin{align}
 \label{eq:BCD}
  C_\BCD(t)=\int_0^\infty dy \frac{e^{-(y+i E_\BC) t}}{2\pi}\Big[&G(E^+_\BC-iy)-G(E_\BC^--i y)\Big]\,,
 \end{align}
 %
 where $E_\BC^{\pm}=\lim_{\varepsilon\rightarrow 0} E_\BC\pm \varepsilon$, and the definition of $G(z)$ must be adapted depending on the value of $z$.
 \end{itemize}

Even though each contribution will be calculated numerically, the separation in different terms gives valuable information about the underlying contribution dominating in each parameter regime. Moreover, in some situations these formulas can be used to obtain asymptotic scalings at long times or perturbative couplings, that is, $g\ll J$.

\subsection{Numerical techniques}

The most straightforward method is to solve directly the dynamics of the total system+bath Hamiltonian, that is, $e^{-i H t}\ket{\Psi_0}$, by trotterizing the time and using split methods. This means that we apply the evolution of the bath, $H_B$, in $\kk$-space, whereas the one of $H_\mathrm{int}+H_S$ in position space, as we explained in Ref.~\cite{gonzaleztudela17b}. 

Another option, very useful for 3D models, is to work in the space of frequencies by discretizing $\omega(\kk)$ in steps of $d\omega$. Like this, one can group the evolution of the bath Hamiltonian in different isofrequency subspaces~\cite{gonzaleztudela17b}:
\begin{align}
H_B=\sum_{n=1}^{N_\omega}\omega_n \sum_{\kk=\kk(\omega_n)} \hat{a}^\dagger_{\kk}\hat{a}_{\kk}=\sum_{n=1}^{N_\omega}\omega_n \sum_{\alpha=1}^{\mathcal{N}(\omega_{n})} \hat{a}^\dagger_{\omega_n,\alpha}\hat{a}_{\omega_n,\alpha}\,.
\end{align}
where $N_\omega$ is the total number of frequency modes considered, i.e., $N_\omega=\frac{\mathrm{max}\omega(\kk)-\mathrm{min}\omega(\kk)}{d\omega}$, and $\omega_n=\mathrm{min}\omega(\kk)+n d\omega$.  The index $\alpha$ of the frequency creation/destruction operators $\hat{a}^\dagger_{\omega_n,\alpha}/\hat{a}_{\omega_n,\alpha}$ denotes the degeneracy of the modes at $\omega_n$. This index runs from $1$ to $\mathcal{N}(\omega_n)$, where $\mathcal{N}(\omega_n)$ is the number of modes at the frequency step $\omega_n$. 

To illustrate how it simplifies the evolution, let us consider a single QE coupled to the bath, and choose the convention where the index $\alpha=1$ corresponds to the symmetric combination of the $\kk$ modes, which is the one in this case that couples the QE dipole transition. With this choice, the interaction Hamiltonian reads:
\begin{align}
 H_\intt&=\frac{g}{N^{3/2}}\sum_\kk\left( \hat{a}_\kk^\dagger \sigma_{eg}+\mathrm{h.c.}\right)=\frac{g}{N^{3/2}}\sum_n \sqrt{\mathcal{N}(\omega_n)}\left( \hat{a}_{\omega_n,1}^\dagger \sigma_{eg}+\mathrm{h.c.}\right)
\end{align}

As the interaction Hamiltonian only couples $\sigma_{eg}$ to the $\alpha=1$ indices, for each $\omega_n$ one can neglect the dynamics of the $\mathcal{N}(\omega_n)-1$ modes that are uncoupled from the QE. With this method, we have empirically observed one can obtain the dynamics up to a given time, $T\lesssim 1/d\omega$. This means, the longer the time we want to explore, the more frequency modes we need to accurately describe the dynamics.

\subsection{Dynamics in the Markovian regime.}

To conclude, it is instructive to remind the reader the predictions for the dynamics obtained in the Markovian/perturbative regime for the problems that we consider along the manuscript. First, let us mention that the Markovian results are easily recovered in our formalism just by replacing $\Sigma(E+i0^+)\approx \Sigma(\Delta+i0^+)$ in Eqs.~\ref{eq:self}-\ref{eq:inv}, and assuming that the only contribution to the integral of Eq.~\ref{eq:inv} is coming from a single pole: $z=\Delta+\Sigma(\Delta+i0^+)$. This approximation predicts:
\begin{itemize}

 \item For a single QE initially excited, the probability amplitude of the excited state is given by: 
 \begin{align}
   C_e(t)\approx e^{-i\left(\Delta+\delta\omega_M-i\frac{\Gamma_M}{2}\right)t}\,,
 \end{align}
 where $\delta\omega_M$ and $\frac{\Gamma_M}{2}$ are the real and imaginary parts of the self energy $\Sigma_e(\Delta+i0^+)$, given by the formula:
 \begin{align}
  \Sigma_e(\Delta+i0^+)=\frac{g^2}{(2\pi)^3}\iiint d^3\kk \Big[& \mathrm{P.V.}\frac{1}{\Delta-\omega(\kk)}-i\pi\delta(\Delta-\omega(\kk))\Big]\,.
 \end{align}

 Notice, this approximation agrees with the prediction of the Fermi's Golden Rule, which predicts an exponential relaxation of the population $|C_e(t)|^2=e^{-\Gamma_\mathrm{FGR} t}$, with a decay rate given by: $\Gamma_\mathrm{FGR}=2\pi g^2 D(\Delta)$, where $D(\Delta)$ is the density of states of the bath, which in the thermodynamic limit reads:
 \begin{align}
 \label{eq:dos}
  D(E)&=\frac{1}{(2\pi)^3}\iiint \delta(E-\omega(\kk))=-\frac{\mathrm{Im}\Sigma_e(E+i0^+)}{g^2 \pi}\,.
 \end{align}

 \item When more than one QE are coupled to the bath, there are two different situations. When we start in an initial state which couples only to a collective bath mode $A_\alpha$, the dynamics is analogous to the single QE case, but with a modified self-energy $\Sigma_\alpha(\Delta+i0^+)$ which now will contain extra terms. This will be the case of two QEs initialized in the symmetric/antisymmetric superposition because these two states coupled to orthogonal bath modes, as long as $\omega(\kk)=\omega(-\kk)$~\cite{gonzaleztudela17b}. In that case, the dynamics of these states is governed by a modified self-energy, which now depends on the relative distance position between the QEs, $\nn_{12}$, as follows $\Sigma_{\pm}(z;\nn_{12})=\Sigma_e(z)\pm \Sigma_{12}(z;\nn_{12})$ for the symmetric/antisymmetric QE superposition, with:
\begin{align}
 \label{eq:self12}
 \Sigma_{12}(z;\nn_{12})=\frac{g^2}{(2\pi)^3}\iiint d^3\kk \frac{e^{i\kk\cdot\nn_{12}}}{z-\omega(\kk)}\,.
\end{align}

The perturbative prediction is the same than for a single QE, but with $\Sigma_{\pm}(\Delta+i0^+)=J_{M,\pm}-i\frac{\Gamma_{M,\pm}}{2}$.
 
 \item The other situation occurs when we start in a state that couples to many bath modes, e.g., for the situation with two QEs with one of them initially excited. In that case, one needs to consider the coupling to both the symmetric/antisymmetric bath modes~\cite{gonzaleztudela17b}. For example, when only the first is excited, the dynamics of the two QEs reads:
\begin{align}
 \label{eqHC:c12}
 |C_{1,2}(t)|^2&\approx \frac{1}{4}\Big[ \pm 2 e^{-\frac{\Gamma_{M,+}+\Gamma_{M,-}}{2}t}\cos((J_{M,+}-J_{M,-})t) +e^{-\Gamma_{M,+} t}+e^{-\Gamma_{M,-} t}\Big]\,,
\end{align}
which predicts a coherent exchange of excitations at rate $J_{M,+}-J_{M,-}$, exponentially damped by $\Gamma_{M,\pm}$. 

\end{itemize}

\section{Bath properties: geometries and density of states \label{sec:bath}}

\begin{figure*}
\centering
\includegraphics[width=0.99\textwidth]{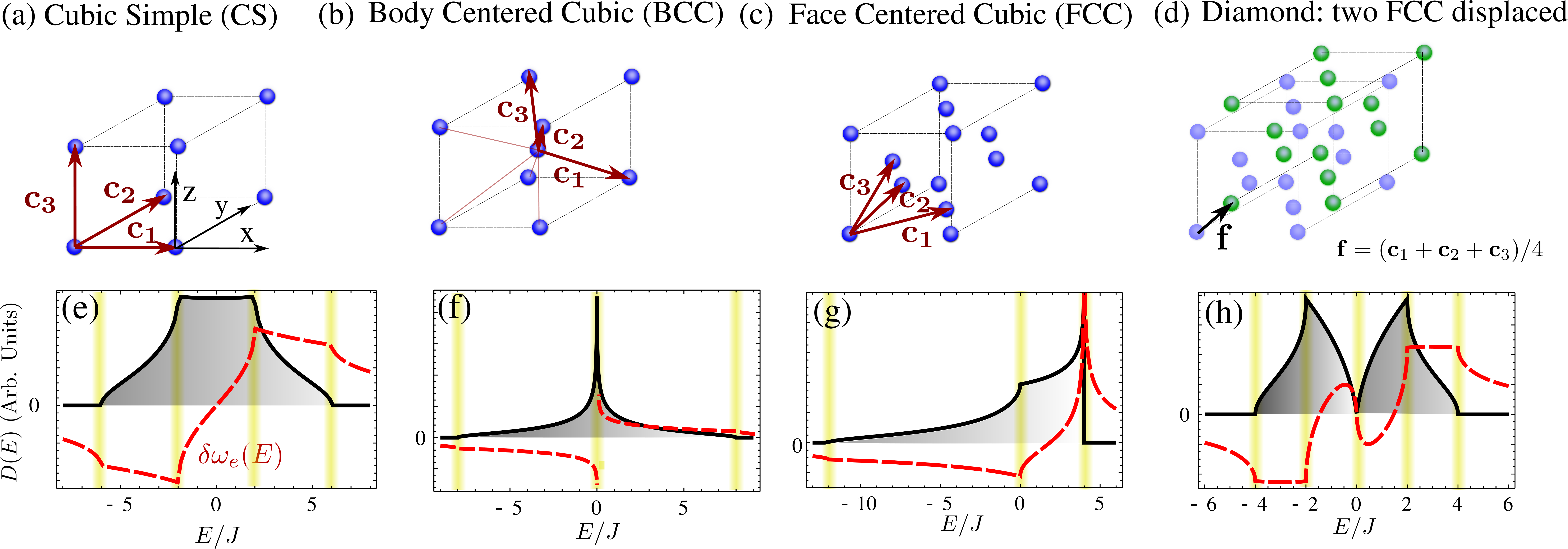}
\caption{(a-d) Bath geometries and primitive vectors $\cc_i$ for the 3D structured baths considered CS, BCC, FCC and diamond lattice respectively. (e-h) Corresponding density of states $D(E)$ (proportional to $\Gamma_M(E)$ as shown in Eq.~\ref{eq:dos}) of the different baths (solid black). For completeness, we also plot $\delta\omega_e(E)$ using the analytical formulas of $\Sigma_e(E)$ given in Sections~\ref{sec:CS}-\ref{sec:diamond}. In shaded yellow, we highlight the regions where non-perturbative dynamics is expected.}
\label{fig:3}
\end{figure*}

After having introduced the general setup, implementation, and the techniques to deal with these problems, let us present the reservoirs that we consider along this manuscript. They will be mainly nearest neighbours tight-binding models with different geometries well-known in the condensed matter context~\cite{ashcroft76a}. However, given the importance that they will have for the discussion of the results, we present in this Section their primitive vectors, energy dispersions, and their corresponding density of states, $D(E)$, in the thermodynamic limit. The latter is particularly relevant since it gives us hints on where non-perturbative dynamics is expected due to the appearance of non-analytical behaviour. We have chosen four different models with qualitative differences in their density of states, as we show in Fig.~\ref{fig:3}, and therefore, where we expect to obtain different dynamical features.

\subsection{Cubic Simple (CS) lattices}

We start describing the most simple 3D structured bath, namely, the one given by a CS geometry depicted in Fig.~\ref{fig:3}(a). In this case, considering the length of the natural cube $L$ as the unit of length, the primitive vectors spanning the lattice are: $\cc_{1,2,3}=\hat{\mathbf{e}}_{x,y,z}$. Each lattice site has 6 nearest neighbours at positions: $\pm \cc_{1,2,3}$. This can be shown to lead to an energy dispersion:
\begin{equation}
 \label{eq:omCS}
 \omega_\CS(\kk)=-2J\left[\cos(k_1)+\cos(k_2)+\cos(k_3)\right]\,,
\end{equation}
where $\kk=(k_1,k_2,k_3)$ is the momenta given with respect to the primitive lattice coordinates, $\dd_i$, which in the CS lattice are: $\dd_{1,2,3}=2\pi \hat{\mathbf{e}}_{x,y,z}$. Notice, the reciprocal space for a CS lattice is another CS lattice. It can be easily shown that this band expands from $[-6J,6J]$, and its corresponding density of states, shown in Fig.~\ref{fig:3}(e), has several spectral regions which are candidates to display non-perturbative dynamics (in shaded yellow in the figure). For example, at the band-edges the density of states is continuous but with discontinuous derivative. It is easy to show that around the upper/lower edge the density of states is isotropic, e.g., in the lower one $\omega_\CS(\kk)/J\approx -6+ |\kk|^2$, which leads to:
\begin{equation}
 \label{eq:dosCS3D}
 D_\CS(E)\approx \frac{1}{4\pi^2 J}\sqrt{\frac{E+6J}{J}}
\end{equation}
for $E\gtrsim-6J$. This is a feature of most 3D band-edges, which is also captured by simplified isotropic models considered in the literature~\cite{devega08a,navarretebenlloch11a}. However, the CS lattice already shows some distinctive features not captured by isotropic models, namely, at $E=\pm 2J$, $D(E)$ has again a discontinuous derivative but with a finite value of the density of states. As we show in Section~\ref{sec:CS}, this will lead to non-Markovian relaxation dynamics different from other types of reservoirs.

\subsection{Body-Centered Cubic (BCC) lattices}

The next lattice that we consider is the BCC lattice, depicted in Fig.~\ref{fig:3}(b), which is characterized by having an extra bosonic mode in the center of the unit cube. This is also a simple Bravais lattice, which can be expanded by choosing the following primitive vectors: $\cc_{1}=(\hat{\mathbf{e}}_x+\hat{\mathbf{e}}_y-\hat{\mathbf{e}}_z)/2$, $\cc_{2}=(\hat{\mathbf{e}}_x-\hat{\mathbf{e}}_y+\hat{\mathbf{e}}_z)/2$, and $\cc_{3}=(-\hat{\mathbf{e}}_x+\hat{\mathbf{e}}_y+\hat{\mathbf{e}}_z)/2$. Each bosonic mode has 8 nearest neighbours, which position written in the basis of primitive vectors is given by: $(\pm 1,0,0),(0,\pm 1,0), (0,0,\pm 1), (\pm 1,\pm 1,\pm 1)$. In this case, the reciprocal vectors are given by: $\dd_{1}=2\pi(\hat{\mathbf{e}}_x+\hat{\mathbf{e}}_y)$, $\dd_{2}=2\pi(\hat{\mathbf{e}}_x+\hat{\mathbf{e}}_z)$, and $\dd_{3}=2\pi(\hat{\mathbf{e}}_y+\hat{\mathbf{e}}_z)$. Diagonalizing the bath Hamiltonian in the momentum space, one arrives to the following energy band dispersion:
\begin{align}
 \label{eq:ombcc}
 \omega_\BCC(\kk)=\omega_\CS(\kk)-2J\cos(k_1+k_2+k_3)\,.
\end{align}

This band expands from $[-8J,8J]$, as shown in the corresponding density of states of Fig.~\ref{fig:3}(f). Apart from the standard scaling $D(E)\propto \sqrt{E}$ at the band-edges, due to the isotropic character of the dispersion at these points, this bath has another prominent feature at $E\approx 0$, where $D(E)$ diverges. In Section~\ref{sec:BCC}, we will prove that this is a square logarithmic divergence, $D(E)\propto \ln(E)^2$, which translates into prominent features in the QE individual and collective dynamics.

\subsection{Face-Centered Cubic (FCC) lattices}

The other simple Bravais lattice that we consider is the FCC lattice, depicted in Fig~\ref{fig:3}(c), which is obtained when there is one extra bosonic mode in the center of each side of the unit cube. The three primitive vectors that expand this lattice are: $\cc_{1}=(\hat{\mathbf{e}}_x+\hat{\mathbf{e}}_y)/2$, $\cc_{2}=(\hat{\mathbf{e}}_x+\hat{\mathbf{e}}_z)/2$, and $\cc_{3}=(\hat{\mathbf{e}}_y+\hat{\mathbf{e}}_z)/2$, whereas the three reciprocal ones are: $\dd_{1}=2\pi(\hat{\mathbf{e}}_x+\hat{\mathbf{e}}_y-\hat{\mathbf{e}}_z)$, $\dd_{2}=2\pi(\hat{\mathbf{e}}_x-\hat{\mathbf{e}}_y+\hat{\mathbf{e}}_z)$, and $\dd_{3}=2\pi(-\hat{\mathbf{e}}_x+\hat{\mathbf{e}}_y+\hat{\mathbf{e}}_z)$. Notice that the reciprocal space of the FCC lattice is a BCC one and viceversa. In real space, each bosonic mode of the FCC lattice interacts with 12 nearest neighbours: $(\pm 1,0,0),(\pm 0,\pm 1,0),(0,0,\pm 1)$, $(\pm 1,\mp 1,0),(\pm 1,0,\mp 1), (0,\pm 1,\mp 1)$, written in the coordinates of the primitive lattice vectors. With 
these neighbours, the energy dispersion of this lattice reads:
\begin{align}
\label{eq:omFCC}
 \omega_\FCC(\kk)&=\omega_\CS(\kk)-2J\Big[\cos(k_1-k_2)+\cos(k_2-k_3)+\cos(k_1-k_3)\Big)\,.
\end{align}

The corresponding density of states of this bath, plotted in Fig.~\ref{fig:3}(g), expands from $[-12J, 4J]$ and show remarkable qualitative differences with respect to the other 3D baths. First, the upper/lower band edges have very different features, unlike the other type of reservoirs. Whereas the lower edge has the typical $D(E)\propto \sqrt{E}$ scaling of 3D reservoirs, the upper edge has a divergence, which we will show in Section~\ref{sec:FCC} to scale as $\propto \ln(4J-E)^2$, for $E \lesssim 4J$. Apart from enhancing the decay rates at this edge, this divergence has important consequences in the emergence of robust atom-photon bound states~\cite{john90a,kurizki90a}.

\subsection{Diamond lattice}

Finally, we consider a bath geometry which cannot be described as a simple Bravais lattice, namely, the diamond lattice. As we plot in Fig.~\ref{fig:3}(d), this lattice can be constructed by two interspersed FCC lattices, whose modes are denoted by $a_\nn,b_\nn$, displaced by a vector $\mathbf{f}=1/4(1,1,1)$. To calculate its corresponding band structure, the relevant information is that each bosonic mode in the $A$ lattice interacts with four nearest neighbours from the $B$ lattice at sites: $(0,0,0), (1,0,0), (0,1,0),(0,0,1)$, and vice-versa. This results into a bath Hamiltonian written in momentum space:
\begin{equation}
 H_{B}=\sum_\kk(f(\kk) \hat{a}^\dagger_\kk \hat{b}_\kk+\hc)\,,
\end{equation}
where $f(\kk)=1+e^{i k_1}+e^{i k_2}+e^{i k_3}=|f(\kk)|e^{i\phi(\kk)}$ is the coupling between the $A/B$ lattices in $\kk$-space. The Hamiltonian can be easily diagonalized by introducing the following operators:
\begin{equation}
 \label{eq:ul}
 \hat{u}_\kk/\hat{l}_\kk=\frac{1}{\sqrt{2}}\left(\hat{a}_\kk\pm e^{i\phi(\kk)}\hat{b}_\kk\right)
\end{equation}
which represent the annhiliation operator of the upper/lower band modes. The resulting eigenenergies, $\omega_{u/l}(\kk)=\pm |f(\kk)|$, have a close connection with the energy dispersion of the FCC lattice:
\begin{equation}
 \label{eq:ulene}
 |f(\kk)|^2=4J-\omega_\FCC(\kk)\,.
\end{equation}

From here, it is very easy to read that the $\kk$-points that give rise to the upper edge of the FCC lattice will be the ones where the upper/lower band of the diamond lattice touch, since $\omega_{u,l}(\kk)\equiv 0$ at these points. The global upper/band edges at $E=\pm 4J$ have the characteristic $D(E)\propto \sqrt{E}$ of isotropic 3D band dispersions. In the middle of each band, at $E=\pm 2J$ there are some kinks at the density of states, similar to the ones obtained for the CS and FCC lattice. The main qualitative difference here is the appearance of a singular band gap point at the position where the two bands touch, i.e., $E=0$. This is the 3D analogue of the Dirac cone in 2D lattices~\cite{castroneto09a}, and it will be the source of qualitatively new features in the QE dynamics.

After having explored how the structure of the different baths give rise to non-analytical features in their corresponding density of states, in the next sections we study the exact individual and collective QE dynamics emerging from them. We will emphasize the features that are different from reservoirs of lower dimensions or simplified descriptions to the problem.

\section{Quantum dynamics in cubic-simple baths: long-time reversible dynamics  \label{sec:CS}}

In this Section we analyze the QE dynamics emerging from CS baths. Since we use the same procedure to study the different baths, we explain it here explicitly to guide the reader in the discussion:
 \begin{itemize}
 \item We always start by analyzing the relaxation of an initially excited QE. We first give the analytical expression of the self-energy $\Sigma_e(z)$. Then, we show how to analytically continue to the whole complex plane to perform the integral of Eq.~\ref{eq:inv} distinguishing the different contributions. 
 
 \item Once we have analyzed the exact dynamics of $C_e(t)$, we study its emission into the bath and spot the parameters to observe interesting collective phenomena such as super/subradiance or coherent exchange of excitations. Then, we analyze the exact dynamics of several QEs interacting with the bath focusing only on those regions (if any).

 \end{itemize}

\subsection{Single QE}

By using the fact $\omega(\kk)=\omega(-\kk)$, the single QE self-energy of the CS bath can be written as:
\begin{equation}
 \label{eq:CSbath1}
 \Sigma_{e,\CS}(z)=\frac{g^2}{\pi^3}\iiint_0^\pi \frac{d^3\kk}{z-\omega_\CS(\kk)}\,.
\end{equation}

An analytical expression of this function can be found~\cite{guttmann10a} in terms of elliptic integrals~\cite{abramowitz66a}, which reads:
\begin{equation}
 \label{eq:CSbath2}
 \Sigma_{e,\CS}(z)=\frac{4 g^2}{\pi^2 z}\frac{1-9\xi^4(z)}{(1-\xi(z))^3(1+3\xi(z))}\KK\left[m(z)\right]^2\,,
\end{equation}
where the functions $\xi(z)$ and $m(z)$ read:
\begin{align}
 \label{eq:aux}
 \xi(z)&=\frac{\sqrt{1-\sqrt{1-\frac{4 J^2}{z^2}}}}{\sqrt{1+\sqrt{1-\frac{36 J^2}{z^2}}}}\,,\\ \label{eq:aux2}
 m(z)&=\frac{16\xi^3}{(1-\xi(z))^3(1+3\xi(z))}\,,
\end{align}
and where $\KK(m)$ is the complete elliptic integral of the first kind:
\begin{align}
 \label{eq:ellipticK}
 \KK[m]=\int_0^{\pi/2}\frac{d\theta}{\sqrt{1-m\sin^2(\theta)}}\,.
\end{align}

Note we have defined the elliptic integral in terms of the square of the elliptic modulus, $k$, i.e., $m=k^2$~\cite{abramowitz66a}. The complete elliptic integral is real as long as $m<1$ and has a branch cut along $\mathrm{Re}m\in[-1,\infty)$. Evaluating this function above the real axis: $\Sigma_e(E+i0^+)=\delta\omega_e(E)-i\frac{\Gamma_e(E)}{2}$, one obtains both the density of states of Fig.~\ref{fig:3}(a) (up to a factor is the same as $\Gamma_e(E)$), and the Lamb-shift $\delta\omega_e(E)$, plotted in dashed red in the same figure.

\begin{figure*}
\centering
\includegraphics[width=0.99\textwidth]{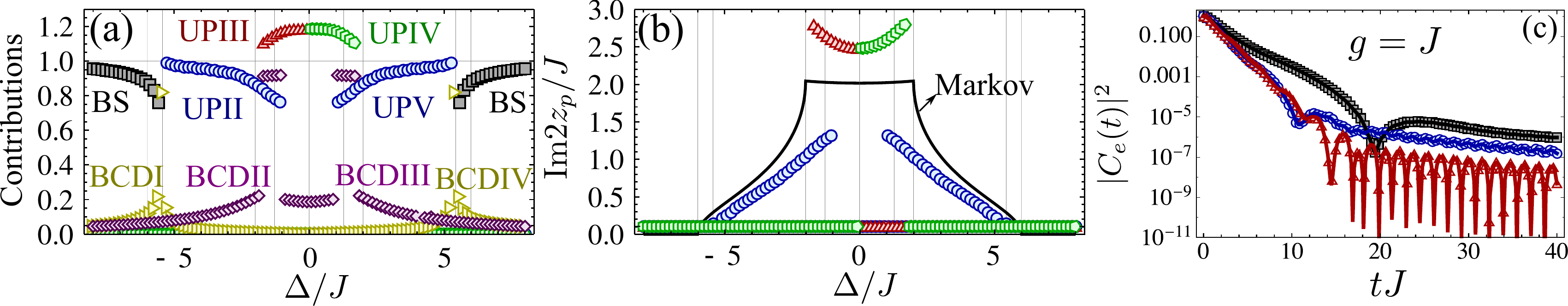}
\caption{(a) Absolute value of the weight of the different contributions of $C_e(0)$  for a CS lattice as a function of $\Delta/J$ for a fixed $g/J=1.5$: upper/lower BS (black squares), UPs of regions II/V (blue spheres) and regions III/IV (red triangles/green pentagons), BCDII+BCDIII (purple rhomboids) and BCDI+BCDIV (yellow triangles). (b) Imaginary part of the UPs for the same parameters as panel (a), compared to the perturbative (Markov) prediction. (c) QE dynamics for $g=J$ and $\Delta/J=-2$ (black), $\Delta/J=-1$ (blue) and $\Delta/J=0$ (red). In solid lines we plot the result from the complex integration of Eq.~\ref{eq:inv}, and in markers a numerical evolution of the complete Hamiltonian using discretized frequency space.}
\label{fig:4}
\end{figure*}

To integrate $C_e(t)$ using Eq.~\ref{eq:inv} and the Residue Theorem, one needs to avoid the non-analytical regions of $\Sigma_{e,\CS}(z)$ by taking detours in the contour of integration. In this particular case, one possibility is to take four detours at energies $E=-6J,-2J,2J,6J$. As we explained in Section~\ref{sec:theory}, when taking these detours it is possible that one needs to analytically continue the function to other Riemann sheets and, therefore, change the expression of $\Sigma_{e,\CS}(z)$. In this case, we will need to define the $\Sigma_{e,\CS}(z)$ in the complex lower plane in six different regions:
 \begin{itemize}
 \item The definition of Eqs.~\ref{eq:CSbath2}-\ref{eq:ellipticK} is valid for regions where $\mathrm{Re}z\in (-\infty,-6J)$ and $[6J,\infty)$, which we label as regions I and VI.
 
  \item  When  $-6J<\mathrm{Re}(z)<-2J$ and $2J<\mathrm{Re}(z)<6J$, which we denote as regions II and V, the $\sqrt{1-36J^2/z^2}$ of $\xi(z)$ becomes complex. Thus, in order to analytically continue to another Riemann sheet here, one needs to change $\sqrt{1-36J^2/z^2}\rightarrow -\sqrt{1-36J^2/z^2}$, in the definition of $\xi(z)$.
 
  \item  When $-2J<\mathrm{Re}(z)<0$/$0<\mathrm{Re}(z)<2J$, which we label as regions III/IV respectively, both square roots $\sqrt{1-36J^2/z^2}$ and $\sqrt{1-4J^2/z^2}$ become complex such that one needs to change the sign of both of them in $\xi(z)$ to go to the other Riemann sheet.
 However, this introduces one complication: with the new definition of $\xi(z)$, the argument of the elliptic integral crosses its branch cut when $\mathrm{Re}z=0$. Thus, one needs to adapt the definition of $\Sigma_{e}(z)$ for regions III and IV. In particular, if $\mathrm{Im}(m(z))</>0$, then $\Sigma_{e,\CS}(z)$ is defined in the same way as in Eqs.~\ref{eq:CSbath2}, whereas if $\mathrm{Im}(m(z))>/<0$, then $\mathrm{K}[m]\rightarrow \mathrm{K}(m)\mp 2i\mathrm{K}[1-m]$. If one does not change the definition of $\mathrm{K}(m)$, function $\Sigma_{e,\CS}(z)$ will be discontinuous in the detours at $E=\pm 2J$.
 \end{itemize}

With this piecewise definition of $\Sigma_{e,\CS}(z)$, we can integrate $C_e(t)$ using Eq.~\ref{eq:inv} separating its different contributions: BS, UP and the ones coming from the BCDs (see Fig.~\ref{fig:2}(b)). To illustrate the different dynamical regimes that appear in this reservoir, we consider a fixed $g/J$ and plot the (absolute value) of the weight of the different contributions as a function of $\Delta/J$, that we show Fig.~\ref{fig:4}(a). Let us now explain in detail the different regimes that we observe:

\emph{Outside of the band.}
For $|\Delta/J|\gg 6J$, the dynamics is dominated by the BS contribution, i.e., $|R_\BS|\approx 1$ (in solid black in Fig.~\ref{fig:4}(a)). This is expected since there are no bath modes which can lead to the relaxation in the bath and it also happens in lower dimensional baths. As $\Delta/J$ gets closer to the band-edge, the BS contribution starts decreasing until a critical value of $\Delta_\mathrm{crit}$ where it suddenly goes to $0$. This is different from lower dimensional baths, where the BS are more robust since they survive for all values of $\Delta$. The origin of this disappearance of the 3D BS can be traced back to the finite value of $\delta\omega_e(\pm 6 J)$ at the band edge (see Fig.~\ref{fig:3}(a)), compared to the divergent behaviour of $\delta\omega_e(E)$ for 1D and 2D~\cite{gonzaleztudela17b,gonzaleztudela18c}. Intuitively, the bath can only \emph{push} the energy of the BS out of the band up to a critical value, which actually can be calculated to be:
 \begin{equation}
  \label{eq:critdelta}
  \Delta_\mathrm{crit,u/l}=\pm 6J\mp 0.253\frac{g^2}{J}\,.
 \end{equation}
for the upper/lower edge respectively. The dynamics at these band-edges is then dominated by a combination of the BS, and the BCD contributions which gives a subexponential decay of the population. This is a natural feature of 3D isotropic band-edges that was already predicted in Refs.~\cite{john90a,devega08a,navarretebenlloch11a,shi16a}. For this reason, we will not discuss it further and rather focus on the qualitatively new regimes emerging from our structured reservoir.

\begin{figure*}
\centering
\includegraphics[width=0.9\textwidth]{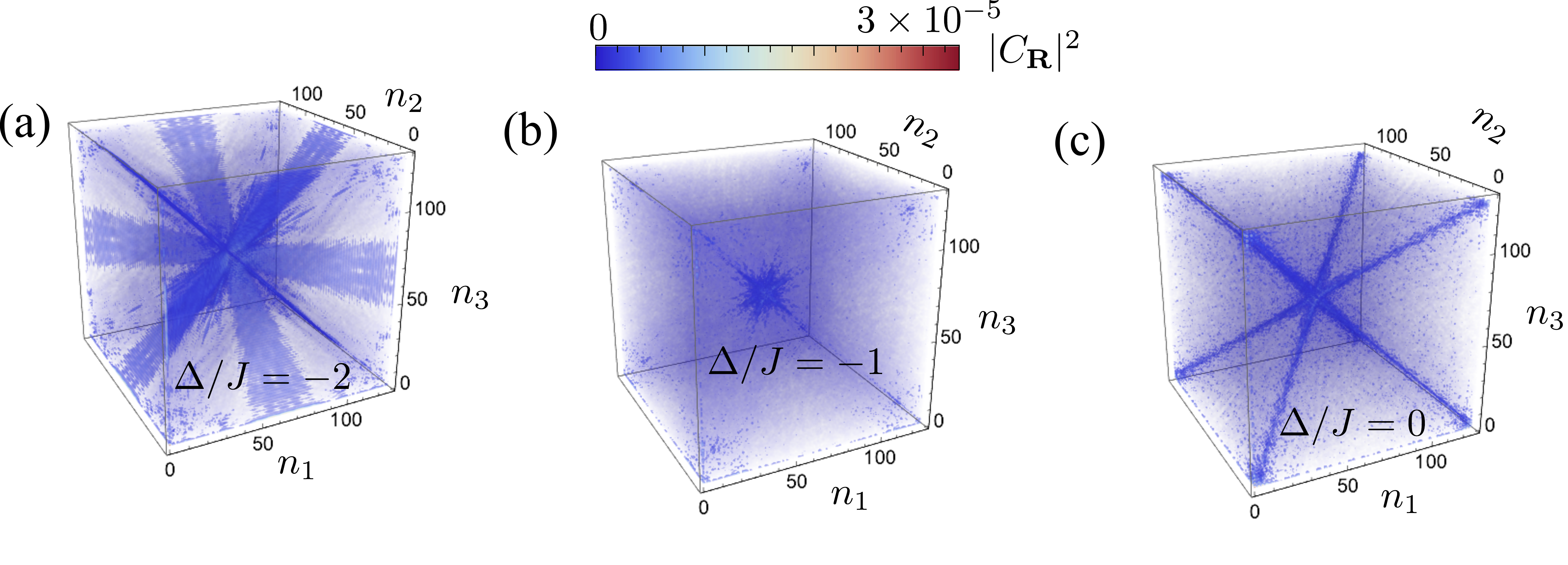}
\caption{(a-c) Bath population in real space, $|C_\RR|^2$, emitted from a single QE coupled to a CS bath with $g/J=0.1$ obtained numerically with a finite bath with linear size $N=2^7$ and for a time $T J=N$}
\label{fig:5}
\end{figure*}

\emph{Inside of the band.}
In the middle of the band the dynamics is generally dominated by the UP contribution, that is, the one due to the contribution of complex poles obtained by using the analytical continuation of $\Sigma_{e,\CS}(z)$ in the different regions like we defined before. As it also occurs for other reservoirs, in the strongly interacting regime ($g\sim J$) the imaginary part of these UPs deviate significantly from the perturbative prediction, as shown in Fig.~\ref{fig:4}(b).

Beyond this non-perturbative renormalization of the lifetimes, two non-trivial spectral regions emerge in the middle of the band at energies $\Delta\approx \pm 2J$, because of non-analytical behaviour of $\Sigma_{e,\CS}(z)$ around them. This behaviour has two consequences: i) there are two extra BCD contributions in the middle of the band (plotted jointly in purple rhomboids in Fig.~\ref{fig:4}(a)); ii) there are two critical spectral regions where two UPs can contribute simultaneously to the dynamics. Although this looks similar to the behaviour emerging from 2D Van Hove singularities~\cite{gonzaleztudela17a,gonzaleztudela17b}, it also shows different features:
 \begin{itemize}

 \item For example, the $\delta\omega_e(E)$ in 2D reservoirs experiences discontinuous jump from positive to negative values, which creates a symmetric region around the 2D divergence where the UP contributions coexist. In the 3D case, the $\delta\omega_e(E)$ is continuous, with a value of $\delta\omega_e(\pm 2J)\approx \mp 0.321 g^2/J$, and only has a discontinuous derivative. Consequently, this makes that while the UP from region II/V can contribute for $|\Delta|<2J$, the UP contribution of regions II and III only survive for $\Delta\in (-2J,2J)$.  Thus, the coexistence regions of the UPs appear in this case at $\sim(-2J,-2J+ 0.321 g^2/J)$ and $(2J- 0.321 g^2/J,2J)$. In these regions, the dynamics shows overdamped oscillations followed by slow relaxation dynamics, as shown in Fig.~\ref{fig:4}(c) for $\Delta=-2J$, similar to the one reported in 2D reservoirs~\cite{gonzaleztudela17b}. The origin of the oscillations is the 
different real part of the UPs giving rise to the dynamics, while the subexponential decay comes from the BCD contribution.
 
 \item Between these two regions, this bath induces a dynamical effect different from other reservoirs. As shown in Fig.~\ref{fig:4}(c), a QE at $\Delta=0$ experiences first an exponential decay, which in the perturbative regime happens at a rate $\Gamma_M(0)\approx 0.89 g^2/J$. However, after this initial relaxation the QE displays coherent oscillations with a very slow decay. These oscillations can be attributed to the interference between the two BCD contribution and they are only weakly attenuated by a power-law decay. Since we have the analytical formulas for $\Sigma_{e,\CS}(z)$ and $\xi(z)$ in Eqs.~\ref{eq:aux} and \ref{eq:CSbath2}, respectively, we can expand the integrand of Eq.~\ref{eq:BCD} around $z=\pm 2J -i y$ for $|y|\ll 1$, which gives the asymptotic limit for $t\rightarrow \infty$ of the BCD contribution. Since $\xi(\pm 2J -i y)\sim a+b\sqrt{y}$, with $a,b\in \mathbb{C}$, this propagates to the integral to arrive to $C_\BCD(t)\propto \int dy \sqrt{y} e^{
- y t}\propto 1/t^{3/2}$. Thus, the non-Markovian 
coherent 
oscillations decay with $|C_e(t)|^2\propto 1/t^3$ in the asymptotic limit.
 
\end{itemize}

To conclude the study of this bath, it is illustrative to characterize how the excitation from a single QE gets emitted into the bath, especially around the plateau between $[-2J,2J]$, which is where the dynamics substantially deviates from the isotropic models. This is what we show in Fig.~\ref{fig:5}, where we plot the bath population in real space, $|C_\RR|^2$ (in primitive coordinates), in a finite bath of size $N=2^7$ after the QE has decayed for a time $T J=N$. We observe that in spite of having similar decay rates (because of the plateau in the density of states), the spatial decay at $E=-2J$ differs significantly from the one at the middle of the band ($E=0$). In particular, we observe that at the kinks of the density of states, $E=\pm 2J$, the excitation gets emitted in stripes around the QE, whereas at $E=0$ it is emitted in 4 directions, namely, at $(1,1,1)$, $(1,1,-1)$, $(1,-1,1)$ and $(-1,1,1)$. 

Both types of emission at $E=2J$ and $0$ give rise to highly anisotropic collective decays when more QEs are coupled to the lattice. However, there seems not to be a perfect subradiant state with many QEs.  Since the other lattice geometries display more interesting collective phenomena, we prefer to stop the discussion of CS bath here, and move to the next considered reservoir.

\section{Quantum dynamics in body-centered-cubic baths: 3D perfect subradiant states  \label{sec:BCC}}

The next bath geometry that we consider is the BCC lattice, whose density of states (see Fig.~\ref{fig:3}(f))) differs significantly from the CS one since it displays a divergence in the middle of the band.

\subsection{Single QE}

 In the continuum limit, the single QE self-energy reads:
\begin{equation}
 \label{eq:selfBCC}
 \Sigma_{e,\BCC}(z)=\frac{g^2}{8\pi^3}\iiint_{-\pi}^\pi \frac{d^3\kk}{z-\omega_\BCC(\kk)}\,,
\end{equation}
where $\omega_\BCC(\kk)$ is given in Eq.~\ref{eq:ombcc}. Notice, it is not yet justified to change $k_i\rightarrow -k_i$ to restrict the range of integration to $[0,\pi]$ because of the term $\cos(k_1+k_2+k_3)$. It is convenient to make a change of variables:
\begin{align}
   \left[ {\begin{array}{c}
   k_1 \\
   k_2 \\
   k_3 
  \end{array} } \right]&=\left[ {\begin{array}{ccc}
   1 & 1 & -1 \\
   1 & -1 & 1 \\
  - 1 & 1 & 1
  \end{array} } \right] \left[ {\begin{array}{c}
   q_1 \\
   q_2 \\
   q_3 
  \end{array} } \right]=U_\BCC \left[ {\begin{array}{c}
   q_1 \\
   q_2 \\
   q_3 
  \end{array} } \right]\,,
\end{align}

In these new variables the dispersion relation factorizes:
\begin{equation}
\omega_\BCC(\qq)=-8J\cos\left(q_1\right)\cos\left(q_2 \right)\cos\left(q_3\right)\,.
\end{equation}

Since $k_{i}\in (-\pi,\pi)$ in the integral, the new integral values run at most from $[-\pi,\pi]$, but actually the integration region is a parallelepiped, $R$, which goes from $[-\pi,-\pi,-\pi]$ to $[\pi,\pi,\pi]$. The self-energy then reads:
\begin{align}
 \label{eq:selfbcc2}
 \Sigma_{e,\BCC}(z)=\frac{g^2}{2\pi^3}\iiint_{R} \frac{d^3\qq}{z+8J\cos\left(q_1\right)\cos\left(q_2 \right)\cos\left(q_3\right)}\,.
\end{align}
where the factor $4$ is coming from the Jacobian of the transformation, i.e., $|\mathrm{Det}(U_\BCC)|$. One can check that one can transport the parallelepiped into two cubes in $[0,\pi]\times [0,\pi]\times [0,\pi]$ (and $[-\pi,0]\times [-\pi,0]\times [-\pi,0]$), arriving then to the following expression for the self-energy:
\begin{align}
 \label{eq:selfbcc3}
 \Sigma_{e,\BCC}(z)=\frac{g^2}{\pi^3}\iiint_{0}^\pi \frac{d^3\qq}{z+8J\cos\left(q_1\right)\cos\left(q_2 \right)\cos\left(q_3\right)}\,.
\end{align} 

This integral can again be expressed analytically in terms of elliptic integrals~\cite{guttmann10a} as follows:
\begin{align}
 \label{eq:selfbcc}
 \Sigma_{e,\BCC}(z)&=\frac{4 g^2}{\pi^2 z}\left(\KK[m(z)]\right)^2,\,\,m(z)=\frac{1}{2}\left(1-\sqrt{1-\frac{64 J^2}{z^2}}\right)\,.
\end{align}

Evaluating the self-energy slightly above the real axis, i.e, $\Sigma_{e,\BCC}(E+i0^+)$, one recovers the expected perturbative decay rate, $\Gamma_e(E)$, and Lamb-shift, $\delta\omega_e(E)$, plotted in Fig.~\ref{fig:3}(b).  The behaviour of $ \Sigma_{e,\BCC}(z)$ in the middle of the band resembles the one appearing in 2D Van Hove singularities~\cite{vanhove53a,gonzaleztudela17a,galve17a}. However, in this case the $\delta\omega_e(E)$, apart from having discontinuous jump at $E=0$, it diverges.

To be able to separate the different contributions to the dynamics of $C_e(t)$ using Residue Theorem one possibility consists in closing the contour of integration taking three detours at $E=-8J, 0, 8J$ to avoid the non-analyticities of $\Sigma_{e,\BCC}$(z). This defines four different regions in the lower complex plane depending on $\mathrm{Re}(z)\in (-\infty,-8J)$, $(-8J,0)$, $(0,8J)$, $(8J,\infty)$, that we label from I-IV respectively. In regions II-III, the $\sqrt{1-\frac{64 J^2}{z^2}}$ becomes complex, such that in order to go to the other Riemann sheet one must change:  $\sqrt{}\rightarrow -\sqrt{}$. The definition of $\KK(m)$ will be always the one of Eq.~\ref{eq:ellipticK} since $m(z)$ never crosses the branch cut of the elliptic integral in all the integration region.
\begin{figure*}
\centering
\includegraphics[width=0.99\textwidth]{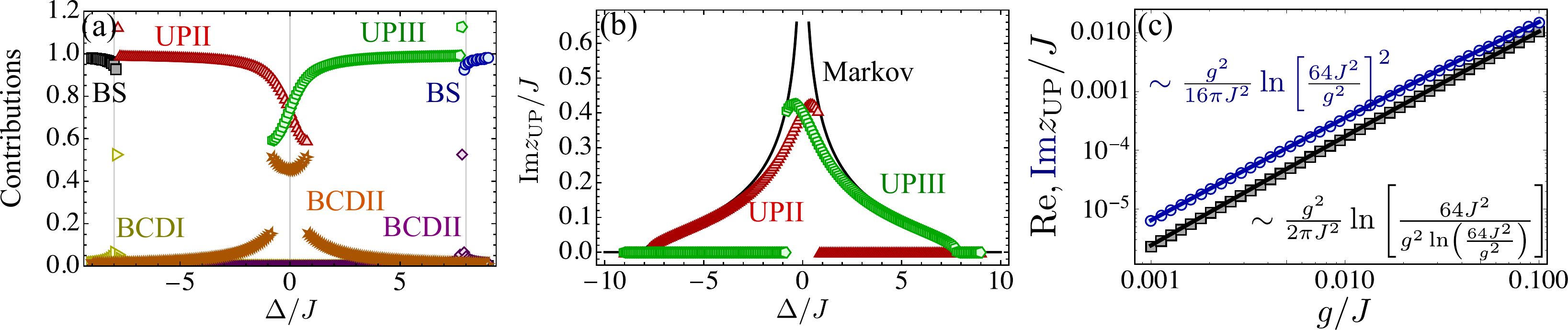}
\caption{(a) Absolute value of the weight of the different contributions of $C_e(0)$ for  a BCC bath as a function of $\Delta/J$ for a fixed $g=J$: upper/lower BS (black squares/blue spheres), UPs of regions II/II(red triangles/green pentagons), BCDI-III (yellow triangles, orange stars, purple rhomboids). (b) Imaginary part of the UPs for the same parameters as panel (a), compared to the perturbative (Markov) prediction. (c) Real and imaginary part of the UP as a function of $g/J$ for $\Delta=0$ obtained numerically by solving the pole equation. In solid lines we plot empirical formulas that approximate the behaviour for the range of $g/J$ considered.}
\label{fig:6}
\end{figure*}

With these prescriptions, we separate the different contributions to $C_{e}(0)$  as a function of $\Delta$ for a fixed $g/J$, as shown in Fig.~\ref{fig:6}(a).  There, we observe that the band-edge effects are similar to the ones obtained in the CS lattice, with a BS contribution that vanishes at a given critical $\Delta$. In the middle of the band, the QE dynamics is mostly governed by the UP, whose lifetimes are also renormalized with respect to the Markov predictions (see Fig.~\ref{fig:6}(b)). Like in the CS lattice, the divergence leads to the coexistence of the UPs due to the divergence of $\Sigma_e(E+i0^+)$ around $E\approx 0$. Moreover, it gives a non-negligible BCD weight at the middle of the band. Since now $\delta\omega_e(E)$ experiences a discontinuous jump from negative to positive around $0$, the coexistence region is now symmetric around the divergence since both the UP in regions II/III can be obtained for $\Delta>/<0$ respectively. To obtain the critical $\Delta$ 
that characterizes the coexistence region of the two UPs, one must check when: $z-\Delta -\Sigma_{e,\mathrm{BCC}}(z)=0$ has still two solutions (using the analytical continuation of the self-energy in the corresponding regions). Expanding $\Sigma_{e,\BCC}(x-iy)$, for $x,y\ll J$, we find:
\begin{align}
 \label{eq:expBCC}
 \mathrm{Re}\Sigma_{e,\mathrm{BCC}}(x-i y)&\approx \frac{g^2}{2\pi J}\ln\left(\frac{64 J}{y}\right)\,,\\
 \mathrm{Im}\Sigma_{e,\mathrm{BCC}}(x-i y)&\approx \frac{g^2}{4\pi^2 J}\left(\pi^2-\ln\left(\frac{y}{64 J}\right)^2\right)\,.
 \end{align}
 
 The complicated shape of the expansion of $\Sigma_{e,\BCC}(z)$ prevented us from finding an analytical solution for the coexistence region of the two BS. Instead, we numerically solved the pole equations for $\Delta=0$, and study the non-perturbative scaling of the real/imaginary part of the UP exactly at the middle of the band (shown in Fig.~\ref{fig:6}(c)). Moreover, in the figure we also give approximated analytical expressions that we found to fit the numerical data in the regions of $g/J$ considered.

\begin{figure}[tb]
\centering
\includegraphics[width=0.99\textwidth]{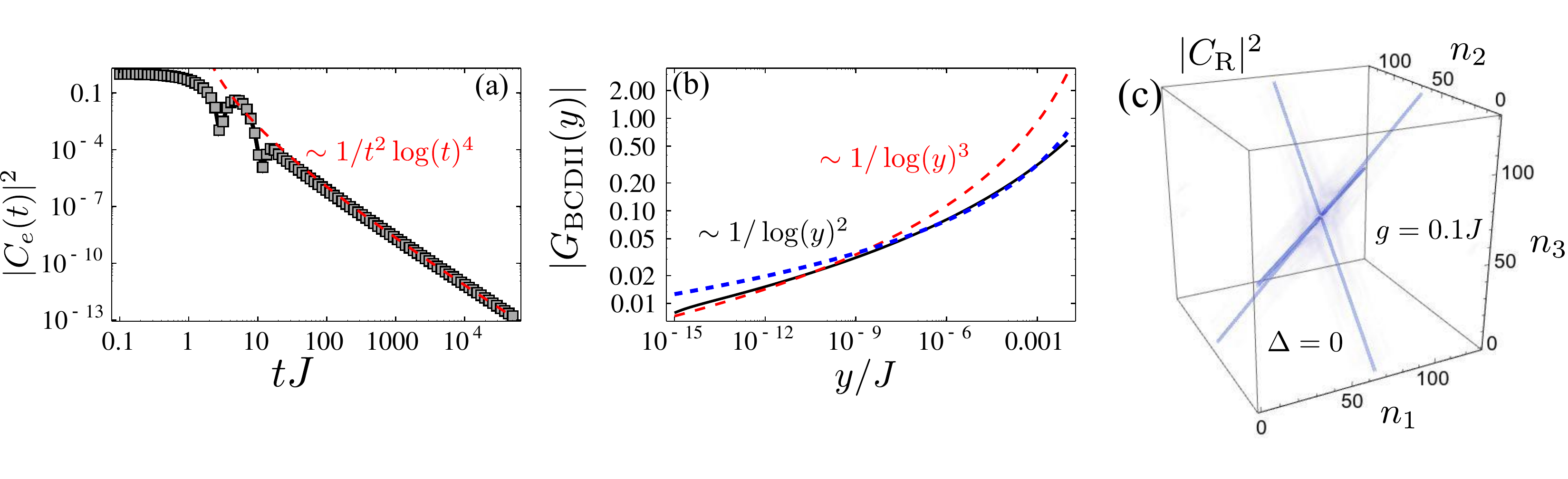}
\caption{(a) $|C_e(t)|^2$ for a single QE coupled to a BCC lattice with $g=J$ and $\Delta=0$ (black markers). In dashed red, we plot the expected asymptotic scaling at long times using Eq.~\ref{eq:asyntoBCC}. (b) Integrand of the BCD contribution at $\Delta=0$ (solid black). As a guide to eye we plot $\ln(y)^{2/3}$ in dashed blue/red to see transitions between the two scalings. (c) Bath probability amplitude, $|C_\RR|^2$, in real space for a QE coupled with $g=0.1J$, $\Delta=0$, after it has emitted into the bath for a time $ TJ=N/2$, with $N=2^7$ the (linear) size of the bath.}
\label{fig:7}
\end{figure}

After having spotted the most interesting spectral region of this reservoir (around $\Delta=0$), let us now study the dynamics at this point. In Fig.~\ref{fig:7}(a), we plot $|C_e(t)|^2$ for a QE coupled to a BCC bath with $g=J$ and $\Delta=0$. One observes two features: first, we observe oscillations in $|C_e(t)|^2$, emerging from the interference of the two UP contributions. These oscillations are overdamped because the imaginary part of the UP is (slightly) larger than their real ones. Secondly, we observe a slow non-exponential relaxation for long times, whose origin is the extra BCD contribution appearing at $\Delta=0$. In principle, one can try to obtain this scaling analytically using Eq.~\ref{eq:BCD}, which reads:
\begin{widetext}
\begin{align}
 C_{\mathrm{BCDII}}(t)&\approx \frac{1}{2\pi}\int_0^\infty G_\mathrm{BCDII}(y) e^{-y t}\,,\\
  G_\mathrm{BCDII}(y)&=\frac{2 \mathrm{Re}\Sigma_{e,\mathrm{BCC}}(0^+-i y)}{(y+\mathrm{Im}\Sigma_{e,\mathrm{BCC}}(0^+-i y))^2+\mathrm{Re}\Sigma_{e,\mathrm{BCC}}(0^+-i y)^2}\,.
\end{align}
\end{widetext}

Since the long-time limit is dominated by the behaviour at $y\ll J$, we can use Eqs.~\ref{eq:expBCC} to obtain $G_\mathrm{BCDII}(y \ll J)$, and then use it to estimate analytically the asymptotic scaling of  $C_{\mathrm{BCDII}}(t)$. As we also show numerically in Fig.~\ref{fig:7}(b), $G_\mathrm{BCDII}(y \ll J)\sim \ln(y)^{-3}$, however, the convergence to this limit is very slow, and for intermediate $y$'s one observes $G_\mathrm{BCDII}(y \ll J)\sim \ln(y)^{-2}$. In both cases, one can use the asymptotic expansion of Fourier integrals in Ref.~\cite{wong78a} to obtain:
\begin{align}
\label{eq:asyntoBCC}
  \lim_{t\rightarrow\infty} C_{\mathrm{MBC}}(t)\sim \frac{1}{t\ln(t)^{2(3)}}\,.
\end{align}

For example, for the ranges of times considered in Fig.~\ref{fig:7}(a) we are probing only the $\ln(y)^{2}$ limit. 

Last, let us plot in Fig.~\ref{fig:7}(c) how a QE spectrally tuned to $\Delta=0$ emits into the bath. There, we observe how the QE excitation spreads into the bath in a very directional fashion, like it also occurred for CS lattices. However, there is an important difference in this case, namely, the emission occurs only in three directions (instead of four),  $(1,1,0),(1,0,1)$ and $(0,1,1)$. As we will see next, this will allow us to find perfect subradiant states when many QEs interact with the bath.

\subsection{Many QEs}

Let us now study collective phenomena when many QEs interact at the same time with a common BCC-type reservoir.  Based on the knowledge of the single QE case, we focus on the case where the QEs are spectrally tuned to the middle of the band, $\Delta=0$, since this is the region where they display a highly anisotropic emission into the bath, which translates into unconventional collective decay terms. Since the goal of the manuscript is to present the most relevant features of each bath geometry, we focus on the emergence of 3D perfect subradiance, something that was thought only to be possible when the QEs are at volume of the order of the wavelength of the QE frequency~\cite{dicke54a} or coupled equally to a cavity mode.

\begin{figure}
\centering
\includegraphics[width=1.\textwidth]{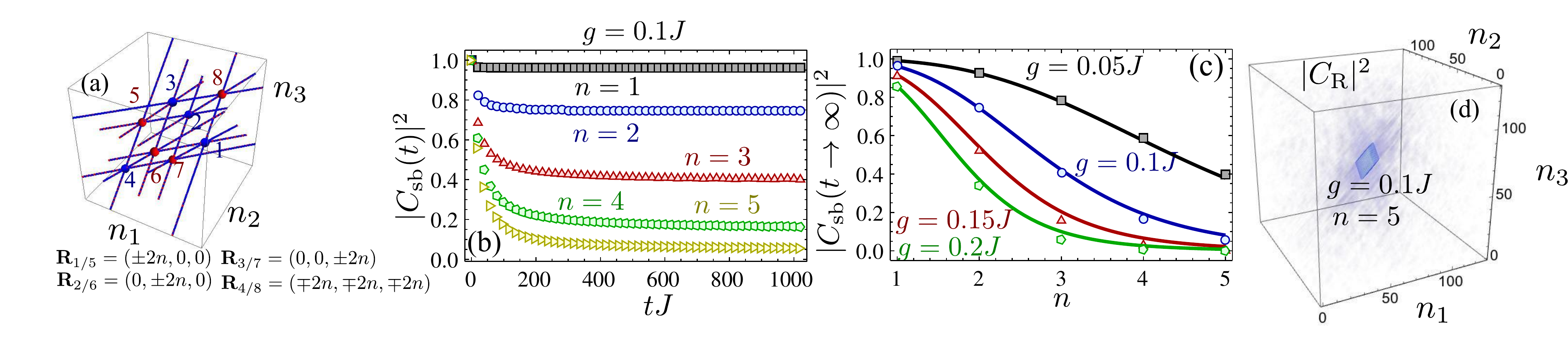}
\caption{(a) Position/phase (plus/minus denoted in colors blue/red respectively) of the 8 QEs to obtain perfect subradiance. Solid lines represent the direction of the emission of each QE. (b) $|C_\mathrm{sb}(t)|^2$, for $g=0.1J$, $N=2^8$, $T J=4 N$, and several distances as depicted in the legend. (c) Long-time population of the subradiant state as a function of $n$ for the same situation of panel (b). Markers are obtained by a numerical simulation of the QE-bath Hamiltonian for $N=2^8$, $TJ=4 N$, and several $g/J$, $n$ as depicted in the legend. Solid lines correspond to analytical results obtained in Eq.~\ref{eq:bcc-res}. (d) Bath population in real space for the situation of panels (b-c) for the distance $n=5$ and after a time $T J= 2^7$.}
\label{fig:8}
\end{figure}

To have perfect subradiance, one needs to be able to cancel perfectly the emission of all QEs at the same time. Since the emission occurs in 3 directions, it can be shown that the minimal configuration to do so is to have $8$ QEs. The QEs must be placed at the positions indicated in Fig.~\ref{fig:8}(a), namely, $(\pm 2n,0,0)$, $(0, \pm 2n,0)$, $(0,0,\pm 2n)$ and $(\pm 2n,\pm 2n, \pm 2n)$, where the parameter $n$ controls the distance between QEs. To prove that a perfect subradiant state emerges in this configuration it is convenient to write $H_\intt$ of Eq.~\ref{eq:Hint} in a collective basis of the QEs. For example, one can use the following rotation between the bare QE basis and a new one:
\begin{align}
 R=\left[ {\begin{array}{cccccccc}
   \frac{1}{\sqrt{8}} &  \frac{1}{\sqrt{8}} &  \frac{1}{\sqrt{8}} & \frac{1}{\sqrt{8}} & -\frac{1}{\sqrt{8}} & -\frac{1}{\sqrt{8}} & -\frac{1}{\sqrt{8}} & -\frac{1}{\sqrt{8}} \\
  \frac{1}{\sqrt{8}} &  \frac{1}{\sqrt{8}} &  \frac{1}{\sqrt{8}} & \frac{1}{\sqrt{8}} & \frac{1}{\sqrt{8}} & \frac{1}{\sqrt{8}} & \frac{1}{\sqrt{8}} & \frac{1}{\sqrt{8}} \\
  \frac{1}{\sqrt{8}} &  \frac{1}{\sqrt{8}} & - \frac{1}{\sqrt{8}} & -\frac{1}{\sqrt{8}} & \frac{1}{\sqrt{8}} & \frac{1}{\sqrt{8}} & -\frac{1}{\sqrt{8}} & -\frac{1}{\sqrt{8}} \\
  \frac{1}{\sqrt{8}} &  -\frac{1}{\sqrt{8}} &  \frac{1}{\sqrt{8}} & -\frac{1}{\sqrt{8}} & \frac{1}{\sqrt{8}} & -\frac{1}{\sqrt{8}} & \frac{1}{\sqrt{8}} & -\frac{1}{\sqrt{8}} \\
  -\frac{1}{\sqrt{8}} &  \frac{1}{\sqrt{8}} &  \frac{1}{\sqrt{8}} &- \frac{1}{\sqrt{8}} & -\frac{1}{\sqrt{8}} & \frac{1}{\sqrt{8}} & \frac{1}{\sqrt{8}} & -\frac{1}{\sqrt{8}} \\
   -\frac{1}{\sqrt{8}} &  \frac{1}{\sqrt{8}} &  \frac{1}{\sqrt{8}} & -\frac{1}{\sqrt{8}} & \frac{1}{\sqrt{8}} & -\frac{1}{\sqrt{8}} & -\frac{1}{\sqrt{8}} & \frac{1}{\sqrt{8}} \\
  0 &  \frac{1}{2} &  -\frac{1}{2} & 0 & 0 & -\frac{1}{2} & \frac{1}{2} &0 \\
  \frac{1}{2} & 0 & 0 &  -\frac{1}{2} & -\frac{1}{2} & 0 & 0 & \frac{1}{2}
  \end{array} } \right] \,
\end{align}

Using this transformation, it is possible to show that the first collective state defined by this transformation, that we denote with annihilation (creation) operators $\sigma_\mathrm{sb}(^\dagger)$, couples only to a bath mode defined by:
\begin{equation}
 A_8^\dagger= \sum_{\qq>0} \frac{1}{8\sin(2 q_1 n)\sin(2 q_2 n)\sin(2 q_3 n)}\sum_{j}a^\dagger_{ U_j \qq}\,,
\end{equation}
where: $U_{j}\rightarrow (q_1,q_2,q_3), (q_1,q_2,-q_3),(q_1,-q_2,q_3),\dots$. One can show that $A_8$ is orthogonal to the bath modes coupled to the other collective QE modes. Thus, if we take as initial state $\ket{\Psi(0)}=\sigma^\dagger_\mathrm{sb}\ket{g}^{\otimes 8}$, the only relevant part of the interaction Hamiltonian reads:
\begin{align}
 H_\mathrm{int}\rightarrow \sum_{\qq>0} \frac{8\sin(2 q_x n)\sin(2 q_y n)\sin(2 q_z n)}{N^{3/2}}\left(A^\dagger_1\sigma_{\mathrm{sb}}+\mathrm{h.c.}\right)\,.
\end{align}

Using this trick, the complexity of the problem reduces to the one of a single QE, but with a collective self-energy which is given by:
\begin{align}
 \label{eq:bcc-self}
 \Sigma_\mathrm{sb}(z)=\frac{64}{(2\pi)^3}2\iiint_0^\pi d^3\qq \frac{\sin^2(2 q_1 n)\sin^2(2 q_2 n)\sin^2(2 q_3 n)}{z+8J\cos(q_1)\cos(q_2)\cos(q_3)}
\end{align}
where the extra factor 2 comes from transporting the parallelepiped into two cubes of size: $[0,\pi]\times [0,\pi]\times [0,\pi]$, like it also happened for the single QE self-energy.

Exploiting the symmetries of the integrand, it is trivial to check that $\Sigma_\mathrm{sb}(0)=0$, which means that there is indeed a real pole in the middle of the band at $E=0$, which therefore does not decay. The only missing thing to prove is that its associated residue is different from zero. Remarkably, since the integrand is separable at $z=0$, this residue can still be calculated analytically, yielding:
\begin{align}
\label{eq:bcc-res}
C_\mathrm{sb}(t\rightarrow\infty)=\frac{1}{1-\partial_z\Sigma_\mathrm{sb}(z)}\Big|_{z=0}=\frac{1}{1+\frac{2g^2 n^3}{J^2}}\,,
\end{align}
which is therefore finite, and close to $1$ if $\frac{2g^2 n^3}{J^2}\ll 1$. In Fig.~\ref{fig:8}(b-d) we certify numerically all these analytical predictions. To start with, in Fig.~\ref{fig:8}(b) we plot the dynamics of $|C_\mathrm{sb}(t)|^2$ for a situation with $g=0.1J$ and several distances parametrized by the number $n$ (see Fig.~\ref{fig:8}(a)) obtained by simulating the full QE-bath Hamiltonian in a system with $N=2^7$ sites. We observe both the perfect subradiant effect for small distances, and the correction imposed by retardation at large ones. This retardation effect appears  due to the finite propagation speed of the excitations in the bath, which makes that the interference can only take place after a certain time which depends on the distance between QEs~\cite{gonzaleztudela17a}. To further explore this retardation effect, we plot $|C_\mathrm{sb}(t\rightarrow\infty)|^2$ for several distances and ratios $g/J$, comparing both the numerical simulations of the full QE-bath Hamiltonian (markers) and 
the analytical prediction of Eq.~\ref{eq:bcc-res}, showing very good agreement between them. Finally, in Fig.~\ref{fig:8}(d), we plot the bath population at a time $tJ=N$ for $g=0.1J$ and $n=5$, showing how the bath emission gets 
trapped between the 8 QEs due to destructive interference between the different decaying paths.

\section{Quantum dynamics in face-centered cubic baths: Anisotropic dipole-dipole interactions  \label{sec:FCC}}

In this section, we consider a bath with a FCC geometry, depicted in Fig.~\ref{fig:3}(c), whose density of states also present qualitative differences with respect to the CS and BCC ones.

\subsection{Single QE} 

The single QE self-energy of the FCC lattice in the continuum limit:
\begin{equation}
\label{eq:selfFCC}
\Sigma_{e,\FCC}(z)=\frac{g^2}{8\pi^3}\iiint_{-\pi}^\pi \frac{d^3\kk}{z-\omega_\FCC(\kk)}\,,
\end{equation}
where $\omega_\FCC(\kk)$ was written in Eq.~\ref{eq:omFCC} in terms of the primitive reciprocal coordinates, $\kk=(k_1,k_2,k_3)$. Like we did for the BCC lattice, it is convenient to make a change of variables:
\begin{align}
   \left[ {\begin{array}{c}
   k_1 \\
   k_2 \\
   k_3 
  \end{array} } \right]=\left[ {\begin{array}{ccc}
   1 & 1 &0 \\
   1 & 0 & 1 \\
  0 & 1 & 1
  \end{array} } \right] \left[ {\begin{array}{c}
   q_1 \\
  q_2 \\
   q_3 
  \end{array} } \right]= U_\FCC \left[ {\begin{array}{c}
   q_1 \\
   q_2 \\
   q_3 
  \end{array} } \right]\,,
\end{align}
under which the dispersion relation transforms to:
\begin{align}
\omega_\FCC(\qq)=-4 J\Big[\cos\left(q_1\right)\cos\left(q_2\right)+\cos\left(q_2\right)\cos\left(q_3\right) +\cos\left(q_1\right)\cos\left(q_3\right)\Big]\,.
\end{align}

The integration region of the self-energy in the new variables, i.e., $q_1=(k_1+k_2-k_3)/2$,\dots, is a polyhedron where each $q$ variable runs at most from $-3\pi/2<q_i<3\pi/2$. Using that the Jacobian of the transformation  $|\mathrm{Det}U_\FCC|=2$, and transporting the polyhedron into the cube $[0,\pi]\times[0,\pi]\times[0,\pi]$, one arrives at
\begin{equation}
 \label{eq:selffcc2}
 \Sigma_{e,\FCC}(z)=\frac{g^2}{\pi^3}\iiint_{0}^{\pi}\frac{d^3\qq}{z-\omega_\FCC(\qq)}\,.
\end{equation}
\begin{figure}
\centering
\includegraphics[width=0.99\textwidth]{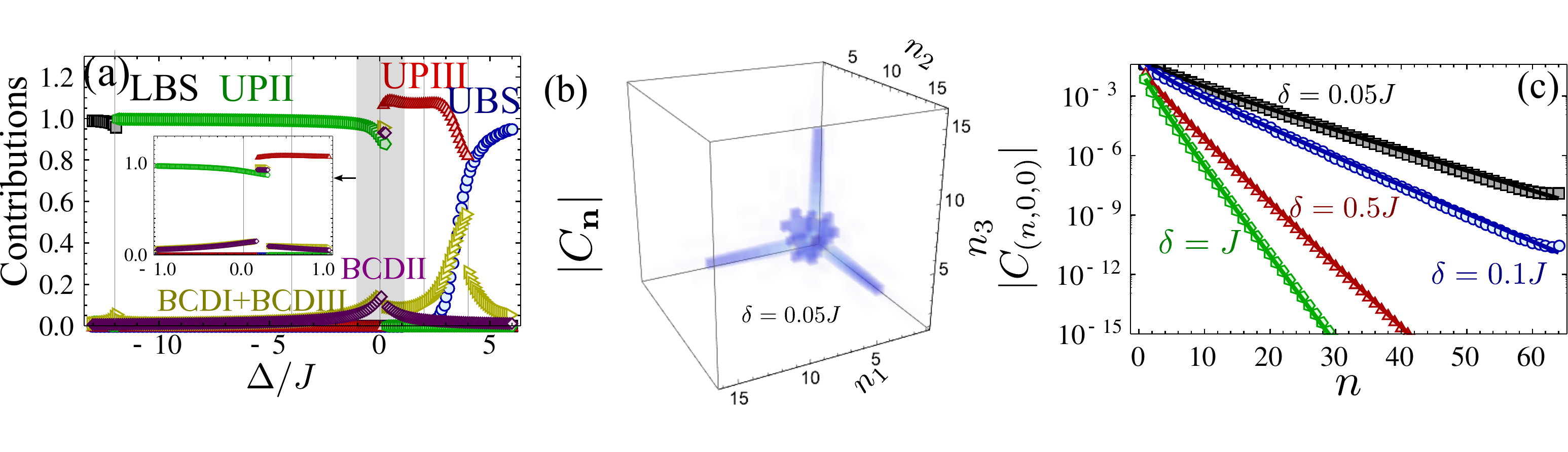}
\caption{(a) Absolute value of the weight of the different contributions of $C_e(0)$ for  a FCC bath as a function of $\Delta/J$ for a fixed $g=J$: lower/upper BS (black squares/blue spheres), UPs of regions II/III(green pentagons/red triangles ), sum of BCDI-III (yellow triangles) and MBC (purple rhomboids). (b) BS wavefunction, $|C_\nn|$, for $\delta=0.05J$, numerically calculated from Eq.~\ref{eq:BSFCC} for system size $N=2^7$. (c) Cut of the BS wavefunction, $|C_\nn|$, along $(n,0,0)$ for several detunings depicted in the legend. The markers correspond to numerical calculation using a finite bath of $N=2^7$, whereas the solid lines represent a numerical approximation given by $0.2 e^{-\sqrt{\delta/J}n}/\sqrt{n\delta}$.}
\label{fig:9}
\end{figure}

Using this transformed expression, it is possible to obtain an analytical solution of the self-energy~\cite{guttmann10a} which reads~\footnote{We correct a small typo appearing in Ref.~\cite{guttmann10a}}:
\begin{align}
 \label{eq:selffcc3}
 \Sigma_{e,\FCC}(z)&=\frac{4 g^2}{\pi^2 z}\frac{(1+3\xi^2)^2}{(1-\xi)^3(1+3\xi)}\left(\KK\left[\frac{16 \xi^3}{(1-\xi)^3(1+3\xi)}\right]\right)^2\,,\\
 \xi(z)&=\frac{-1+\sqrt{1-\frac{4 J}{z}}}{1+\sqrt{1+\frac{12 J}{z}}}\,.
\end{align}
Using this analytical expression we can obtain both the imaginary, $\Gamma_e(E)$, and real part, $\delta\omega_e(E)$, of the self-energy above the real axis, as depicted in Fig.~\ref{fig:3}(g). We observe three spectral regions where $\Sigma_{e,\FCC}(z)$ shows a non-analytical behaviour, namely, $E=-12J, 0, 4J$, and thus they will force us to take a detour in the integration contour to obtain $C_e(t)$. This divides the lower half plane in four regions, where the definition of $\Sigma_{e,\mathrm{FCC}}(z)$ used must be slightly modified depending on the Riemann sheet we want to move in each one:
\begin{itemize}
\item Outside of the band, that is, when $\mathrm{Re}(z)<-12 J$ and $\mathrm{Re}(z)>4J$, that we label as regions I and IV respectively, one can use $\Sigma_{e,\mathrm{FCC}}$ as defined in Eq.~\ref{eq:selffcc3}.

\item Inside of the band one needs to distinguish two regions, from $-12J<\mathrm{Re}(z)<0$ and $0<\mathrm{Re}(z)<4J$, that we denote as regions II and III respectively. In these regions one of the square roots of $\xi(z)$ in Eq.~\ref{eq:selffcc3} becomes complex, and thus, one needs to adapt its definition to go to the adequate Riemann sheet in each integration region. In particular:
\begin{align}
 \label{eq:selffcc3}
 \xi_{\mathrm{II/III}}(z)&=\frac{-1\pm \sqrt{1-\frac{4 J}{z}}}{1\mp\sqrt{1+\frac{12 J}{z}}}\,.
\end{align}
\end{itemize}

With these prescriptions, we can now separate the different contributions to the dynamics of $C_e(t)$ depending on the spectral region, $\Delta$, the QE is coupled to. This is what we summarize in Fig.~\ref{fig:9}(a), where we plot the (absolute value) of the weight of the BS, BCD and UP contributions for a QE coupled with $g=J$. To avoid unnecessary lengthening of the manuscript, we just enumerate common features already appearing in other reservoirs (e.g., CS and BCC) and make an extended discussion on the qualitative new effect emerging in FCC structures:
\begin{itemize}
\item The lower edge at $E=-12J$ shows the typical behaviour of 3D isotropic baths, with a sudden disappearance of the BS contribution.
\item At $E=0$, the density of states is finite, but with a discontinuous derivative. This leads to the coexistence of two UP contributions and an extra BCD contribution in the middle of the band, resulting in non-exponential slow relaxations.
\item The differential feature of this reservoir occurs at its upper band-edge, $E=4J$, where both the real and imaginary part of $\Sigma_{e,\FCC}(z)$ diverge, something very atypical for 3D reservoirs. This divergence leads to two remarkable consequences: i) enhancement of its decay rates as they get closer to the band edge and, ii) the survival of the BS for all the spectral region~\cite{shi16a}.
\end{itemize}

To learn more about this behaviour, we can expand the single QE self-energy around $E\approx 4J+x$, obtaining 
\begin{align}
 \mathrm{Re}\Sigma_{e,\mathrm{FCC}}(4J+ x-i0^+)&=\frac{3 g^2}{16\pi^2 J}\left(-\Theta(-x)\pi^2+\ln\left(\frac{|x|}{64J}\right)^2\right)\,\\
 \mathrm{Im}\Sigma_{e,\mathrm{FCC}}(4J+x-i0^+)&=-\frac{3 g^2}{8\pi J}\ln\left(\frac{|x|}{64J}\right)\Theta(-x)\,,
 \end{align}
where $\Theta(x)$ is the Heaviside function. We have checked numerically (not shown), that these formulas reproduce relatively well the behaviour even for values of $|x|\lesssim J$. Using these formulas for $x>0$, we can obtain an approximated solution for the energy of the upper BS (UBS) for $\Delta=4J$, which reads:
 \begin{align}
  E_{\mathrm{UBS}}\approx 4J +\frac{3 g^2}{4\pi^2 J}W\left(\frac{16\pi J}{g\sqrt{3}}\right)^2\,,
 \end{align}
where $W(x)$ is the so-called Lambert or product-log function, which satisfies $W(x)\approx \ln(1/x)$ for $x\gg 1$ and $W(x)\approx x$ for $x\ll 1$. We have checked (not shown) that this formula reproduces well the behaviour obtained from numerically solving the pole equation at $\Delta=4J$ for a wide range of $g/J$.

Apart from its energy, it is also illustrative to study how the wavefunction of this BS looks like in real space since it will be directly connected to the dipole-dipole interactions when more than one QE is interacting with the bath. The wavefunction can be shown to be given by the following integral:
 \begin{align}
 \label{eq:BSFCC}
  C_\nn&=\frac{1}{N^3}\sum_\kk \frac{e^{i \kk \cdot \nn}}{E_\mathrm{UBS}-\omega_\FCC(\kk)}=\frac{1}{(2\pi)^3}\iiint_{-\pi}^{\pi} d^3\kk \frac{e^{i \kk \cdot \nn}}{E_\mathrm{UBS}-\omega_\FCC(\kk)}\,,
\end{align}
where $\nn=(n_1,n_2,n_3)$ are the primitive coordinates. In Figs.~\ref{fig:9}(b-c), we plot the the numerical evaluation of this wavefunction for a finite system with $N=128$ sites and several detunings with respect to the band-edge: $\delta=E_\mathrm{UBS}-4J$. We highlight two characteristics:
\begin{itemize}
\item The BS is highly anisotropic spreading mainly in the directions $(n,0,0),(0,n,0),(0,0,n)$.
\item To explore the decay along these axis, we note that when $\delta\ll J$, the most relevant $\kk$-points in the integral of Eq.~\ref{eq:BSFCC} are the ones that satisfy $\omega(\kk)=4J$, since they are energetically closer. It can be shown that this equation defines 12 lines in the three-dimensional $\kk$-space, such as $k_1=\pi$ and $k_3-k_2=\pm \pi$, making a closed loop in the integration region. If we are interested in one particular direction, e.g., $(n,0,0)$, the most relevant contribution will come from points where $k_1\approx 0$, e.g., $(0,\pi,\pi)$, since the others will lead to a highly oscillatory integrand. Expanding the dispersion relation around those points we find, e.g., 
\begin{equation}
\label{eq:expFCC}
\omega(q_1,\pi-q_2,\pi-q_3)=4J-J(q_1+q_2+q_3)^2\,,
\end{equation}
for $q_i\ll 1$. Using this expansion, we can make the $q_1$ integral by extending the limits to $(-\pi,\pi)\rightarrow (-\infty,\infty)$ and applying Residue Theorem to arrive at:
\begin{equation}
 |C_{(n,0,0)}|=\frac{e^{-\sqrt{\delta/J}n}}{\sqrt{\delta/J}} F(n)\,,
\end{equation}
which gives us the dependence with $\delta$. The dependence with $n$ is embedded in a function $F(n)$, which unfortunately cannot be calculated using the expansion of Eq.~\ref{eq:expFCC} since gets a divergent integral. Through numerical inspection, we find $F(n)$ can be well fitted by a power-law decay $F(n)=A/n^\beta$, with $A\approx 0.02$ and $\beta\approx 0.5$, as shown in solid lines in Fig.~\ref{fig:9}(c). 
\end{itemize}

These features are very different from the models considered in the literature~\cite{devega08a,navarretebenlloch11a}, where the BSs were always isotropic and with a decay law of the Yukawa type, $\propto e^{-r/\xi}/(r/\xi)$, with $\xi$ being the bound state length and $r$ the distance to the impurity, also inversely proportional of the detuning with respect to the band-edge.

\begin{figure}
\centering
\includegraphics[width=0.8\textwidth]{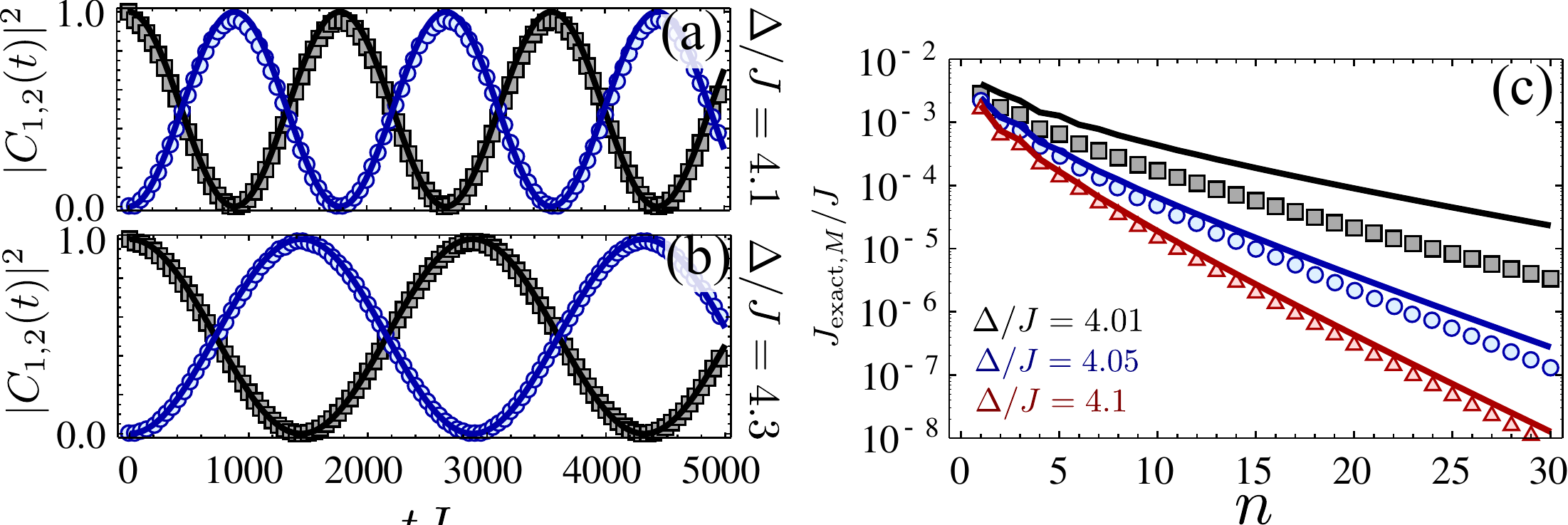}
\caption{(a-b) Dynamics of a pair of QEs separated by $(1,0,0)$ coupled to a FCC lattice for $g=0.1J$, and $\Delta/J=4.1$, $4.3$ respectively. The markers correspond to the results using a numerical simulation of the complete QE-bath Hamiltonian for a bath $N=2^7$, whereas the solid lines are the approximation to cosine/sine function with frequency given by $J_\mathrm{exc}$ as calculated exactly from the pole equation in the symmetric/antisymmetric subspaces. (c) $J_\mathrm{exc}$ (markers) and $J_{M}$ (solid lines) as a function of $n$ for QEs interacting with an FCC bath with $N=2^7$, $g=0.1J$ and several $\Delta/J$ as depicted in the legend. }
\label{fig:10}
\end{figure}

\subsection{Many QEs.}

Using the intuition from other reservoirs~\cite{douglas15a,gonzaleztudela15c,gonzaleztudela18c}, whenever a BS emerges around a QE, it can be used to mediate coherent interactions when many QEs are present in the structure. These interactions will naturally inherit the characteristics of the BS, which in the case of the FCC lattice we have just shown are very distinct from the known behaviours, giving rise to very exotic spin Hamiltonians.

To prove the emergence of these purely coherent dipole-dipole interactions,  we consider a pair of QEs spectrally tuned close to the upper edge, that is, $\Delta\gtrsim 4J$. Their initial state is assumed to be $\ket{\Psi(0)}=\ket{e_1 g_2}$, with one excited and one in the ground state.  In Figs.~\ref{fig:10}(a-b) we show the result of numerical evolution of the complete QE-bath Hamiltonian for two QEs coupled with $g=0.1J$ to an FCC lattice and $\Delta/J=4.1$ and $4.3$, respectively. Their dynamics show indeed that the excitation transfers coherently from one QE to the other with no exponential attenuation. This is what one expects for the Markovian regime from Eqs.~\ref{eq:self12} and \ref{eqHC:c12} in Section~\ref{sec:theory}, which for this particular case lead to:
\begin{align}
\label{eq:FCCpop}
|C_{1 [2]}(t)|^2\approx \cos^2(J_\mathrm{M} t)\left[\sin^2(J_\mathrm{M} t)\right]\,,
\end{align}
where:
\begin{align}
 \label{eq:self12FCC}
 J_\mathrm{M}=\frac{g^2}{N^3}\sum_\kk \frac{e^{i \kk \cdot \nn}}{\Delta-\omega_\FCC(\kk)}\,,
\end{align}
is nothing more than the discrete version of the integral of Eq.~\ref{eq:BSFCC} which gives $C_\nn$. This shows that the dipole-dipole interactions will inherit the exotic properties of the upper BS of the FCC lattice. Moreover, we also notice that in order to accurately capture the frequency of the oscillation for the parameters of Figs.~\ref{fig:10}(a-b), we need to solve the pole equation in the symmetric/antisymmetric subspace exactly, that is, finding:
\begin{equation}
 J_{\pm}-\Delta-\Sigma_{\pm,\FCC}(J_{\pm};\nn)=0
\end{equation}

In the solid lines of Figs.\ref{fig:10}(a-b), we plot the solutions of Eq.~\ref{eq:FCCpop} with a renormalized frequency $J_\mathrm{exact}=(J_-{-}J_{+})/2$ in solid lines showing very good agreement with the results from a simulation of the full QE-bath dynamics. For completeness, in Fig.~\ref{fig:10}(c), we compare $J_\mathrm{exact}$ (markers) versus $J_M$ (lines) for several distances, showing how the Markov approximation becomes more accurate the smaller the distance (or the smaller $g/J$, not shown).

\section{Quantum dynamics in diamond baths: Anisotropic dipole-dipole interactions  \label{sec:diamond}}

In this Section, we explore one example of a 3D two-band model: the diamond lattice. As we explained in Section~\ref{sec:bath}, this bath geometry is formed by two FCC lattices displaced by a vector $1/4(1,1,1)$, as depicted in Fig.~\ref{fig:3}(d). This bath geometry leads to the appearance of a singular band-gap in its density of states, shown in Fig.~\ref{fig:3}(h), which is not present in the other 3D reservoirs considered. Thus, this spectral region will be the main concern of this Section.

\subsection{Single QE}

Using the $H_\intt$ for a two band model, it is easy to show that a QE coupled only to an A (or B) lattice interacts simultaneously with both the upper/lower band. Thus, the single QE self-energy contains now two contributions, which lead to a different expression than for the single band model, that is:
\begin{align}
 \label{eq:selftwoband}
 \Sigma_{e,\mathrm{diam}}(z)=\frac{z g^2}{8\pi^3}\iiint_{-\pi}^\pi \frac{d^3\kk}{z^2-|f(\kk)|^2}\,.
\end{align}
where $|f(\kk)|^2$ was introduced in Eq.~\ref{eq:ulene} in Section~\ref{sec:bath}. Using the same change of variables and integration regions than in the FCC lattice, we can transform the self-energy into:
\begin{align}
 \label{eq:selftwoband2}
 \Sigma_{e,\mathrm{diam}}(z)=\frac{z g^2}{\pi^3}\iiint_{0}^\pi \frac{d^3\qq}{z^2-4J^2\left(1+\sum_{\mathrm{perm}} \cos(q_i)\cos(q_j)\right)}\,,
\end{align}
where $\emph{perm}$ denotes the different permutations of $(q_i,q_j)$ for $i\neq j$. This integral has again an analytical solution expressed in terms of the complete elliptic integral~\cite{guttmann10a}:
\begin{align}
 \label{eq:selftwoband3}
  \Sigma_{e,\mathrm{diam}}(z)=\frac{4g^2}{z\pi^2}\left(\sqrt{4-\frac{16J^2}{z^2}}-\sqrt{1-\frac{16 J^2}{z^2}}\right) \mathrm{K}\left[m(z)\right]^2\,,\\
 m(z)=\frac{1}{2}-\frac{4}{z^2}\sqrt{4-\frac{16 J^2}{z^2}}-\frac{1}{4}\left(2-\frac{16 J^2}{z^2}\right)\sqrt{1-\frac{16 J^2}{z^2}}\,.
\end{align}

\begin{figure}
\centering
\includegraphics[width=1\textwidth]{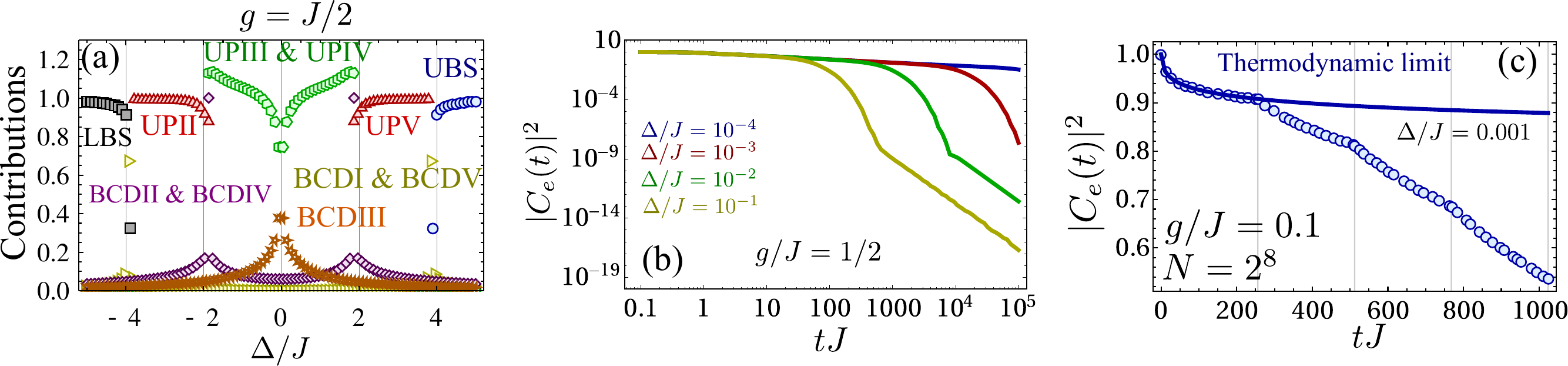}
\caption{(a) Absolute value of the weight of the different contributions of $C_e(0)$ for a diamond bath as a function of $\Delta/J$ for a fixed $g=J/2$: lower/upper BS (black squares/blue spheres), UPs of regions II \& V /III \& IV (red triangles/green pentagons), sum of BCDI-IV (yellow triangles), BCDII \& BCDIV (purple rhomboids) and BCDIII (orange stars). (b) $|C_e(t)|^2$ for a QE coupled with $g=J/2$ for several $\Delta$'s approaching to $\Delta\rightarrow 0$ as detailed in the caption. (c) $|C_e(t)|^2$ for a single QE coupled with $g=0.1J$ and $\Delta=0.001J$ calculated in the thermodynamic limit (solid line) and using a numerical simulation with a finite bath of $N=2^8$ linear size (markers), using $\Delta=0$. }
\label{fig:11}
\end{figure}

Evaluating this expression above the real axis,  $\Sigma_{e,\mathrm{diam}}(E+i0^+)=\delta\omega_e(E)-i\frac{\Gamma_e(E)}{2}$, we have access to both the decay rate $\Gamma_e(E)$ and Lamb-shift, $\delta\omega_e(E)$ that we plot in Fig.~\ref{fig:3}(h). With them, one can easily spot the non-analytical regions where one must take a detour in the contour of integration: the upper/lower band-edges at $E=\pm 4J$, the middle kinks at $E=\pm 2J$ and the singular band-gap at $E=0$. This divides the lower plane in six regions, where one must adapt the definition of $\Sigma_{e,\mathrm{diam}}(z)$ to go to the different Riemann sheets:
\begin{itemize}
\item In the regions $\mathrm{Re}(z)\in (-\infty,-4J)$ and $(4J,\infty)$, denoted by regions I and VI respectively, one can use Eq.~\ref{eq:selftwoband3}.
\item In the regions $\mathrm{Re}(z)\in (-4J,-2J)$ and $(2J,4J)$, denoted as II and V, one must replace: $\sqrt{1-\frac{16 J^2}{z^2}}\rightarrow -\sqrt{1-\frac{16 J^2}{z^2}}$.
\item Finally, when $\mathrm{Re}(z)\in (-2J,0)$ and $(0,2J)$, denoted as regions III and IV, one must change: $\sqrt{4-\frac{16J^2}{z^2}}\rightarrow -\sqrt{4-\frac{16 J^2}{z^2}}$ (as well as $\sqrt{1-\frac{16 J^2}{z^2}}\rightarrow -\sqrt{1-\frac{16 J^2}{z^2}}$).
\end{itemize}

With these prescriptions, we can separate the contributions of $C_e(0)$ of the BSs, BCDs and UPs in the different spectral regions like we did for other reservoirs. This is what we show in Fig.~\ref{fig:11}(a) for a QE coupled to a diamond bath with $g=J/2$. We observe similar phenomena to the other reservoirs at the band edges, $\Delta\approx \pm 4J$, with a sudden disappearance of the BS, and at the kinks, $\Delta\approx \pm 2J$, with coexistence of UPs contributions. The main difference with respect to the other reservoirs occurs at $\Delta\approx 0$, where another BCD must be taken at $E=0$, that gives an extra BCD contribution, denoted as BCDIII and plotted in orange stars in Fig.~\ref{fig:11}(a)~\footnote{The self-energy as written in Eqs.~\ref{eq:selftwoband3} develops numerical instabilities close to the real axis for $|E|\lesssim 10^{-6}$, which is why deliberately avoid these points in Fig.~\ref{fig:11}(a)}. To further understand the behaviour around this point, it is enlightening to expand $\Sigma_{
e,\mathrm{diam}}(E+i0^+)$ around it:
\begin{align}
\label{eq:expdiam}
 \Sigma_{e,\mathrm{diam}}(E+i0^+)&\approx\frac{3g^2 E}{16 J \pi^2}\Big[\ln\left(\frac{64 J^2e^\pi}{E^2}\right)\ln\left(\frac{E^2 e^\pi}{64 J^2}\right)-2\pi i\mathrm{Sgn}(E) \ln\left(\frac{64 J^2}{E^2}\right)\Big]
\end{align}
for $|E|\ll J$.  Since $\Sigma_{e,\mathrm{diam}}(0)=0$, $E=0$ is a solution of the pole equation when $\Delta=0$. However, its derivative $\partial_z\Sigma_{e,\mathrm{diam}}(z)|_{z=0}\rightarrow \infty$ has a logarithmic divergence, and therefore its associated residue in the thermodynamic limit must be zero. This is why as we take $\Delta$ closer to the singular point, the BCDIII contribution becomes more important in Fig.~\ref{fig:11}(a). To certify that, we plot in Fig.~\ref{fig:11}(b) the associated dynamics for the parameters of panel (a), and several detunings approaching $\Delta\rightarrow 0$. We observe that the dynamics is given first by an exponential decay, given by the UP contribution, followed by a subexponential relaxation. Since the imaginary part of the UP also goes to $0$, as predicted by Fermi's Golden Rule and the expansion of Eqs.~\ref{eq:expdiam}, the dynamics becomes slower, and eventually becomes fully dominated by the BCD contribution. A similar behaviour occurred for 2D singular band-
gaps (Dirac points)~\cite{gonzaleztudela18c}, where it was predicted a $1/\ln(t)$ decay for a QE in the thermodynamic limit. For finite $\Delta$, but still close to $0$, the $C_\mathrm{BCDIII}(t)\propto 1/t^2$, which is the power law that we observe in Fig.~\ref{fig:11}(b) for, e.g., $\Delta=0.1J$.

\begin{figure*}
\centering
\includegraphics[width=0.95\textwidth]{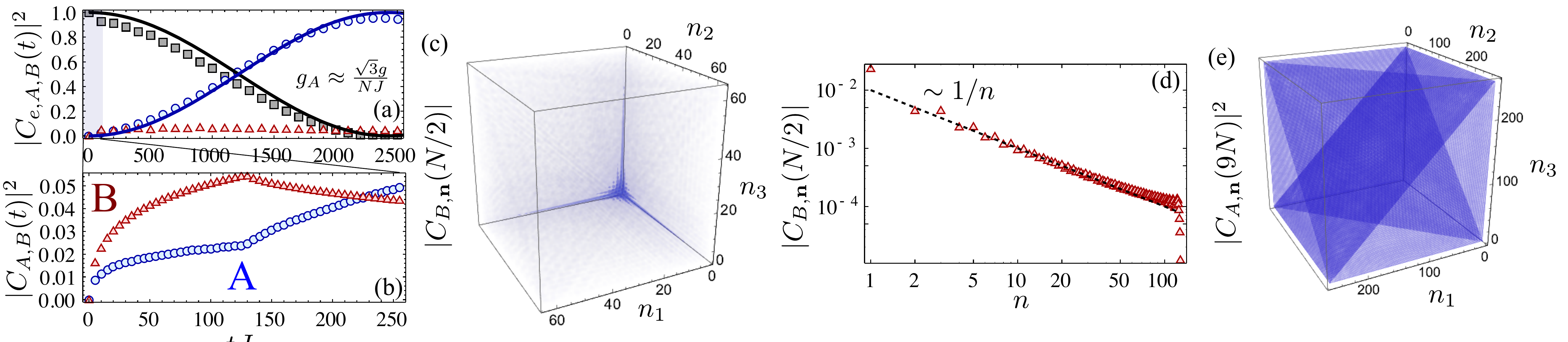}
\caption{(a) $|C_{e,A,B}(t)|^2$ in black squares, blue spheres, red triangles, respectively, for a QE coupled to an A lattice site with $g=0.1J$ to a finite bath of linear size $N=2^8$.  (b) Zoom of panel (a) for short times. (c) Spatial distribution of the B-bath population, $|C_{B,\nn}|$, at a time $tJ=N/2$ for the same situation than panels (a-b). (d) Cut of the spatial distribution of $|C_{B,\nn}|$ along one of the main axes , $\nn=(n,0,0)$, for the same parameters than panel (c). In dashed black, we plot a line $\sim 1/n$ as a guide to eye. (e) Spatial distribution of the A-bath population, $|C_{A,\nn}|^2$, at a time $TJ=9N$, where most of the population is in A-bath as shown in panel (a).}
\label{fig:12}
\end{figure*}

After having spotted similarities with 2D Dirac points, it is instructive to consider the role of finite size effects in the diamond bath, since they were proven to play an important role in the physics emerging at these points at the 2D case. In particular, in 2D it was shown~\cite{gonzaleztudela18c} how for finite systems (with periodic boundary conditions) the spontaneous emission gets quenched because of the emergence of a quasibound photonic state around the impurity. In Fig.~\ref{fig:11}(c), we plot a comparison of $|C_e(t)|^2$ for a QE coupled with $g=0.1J$ and $\Delta/J=0.001$ calculated in the thermodynamic limit using resolvent operator techniques (solid line) and using a finite-bath simulation with periodic boundary conditions (marker). We observe how both calculations agree perfectly well until times $t J\approx N$, being $N$ the linear size of the system. However, in contrast to what happens in 2D singular band-gaps, here the dynamics accelerates instead of quenching. Moreover, we observe that 
this acceleration occurs for times proportional to $N$.

To further investigate this behaviour we make a longer time simulation of $|C_e(t)|^2$ for a QE coupled to the A lattice, that we plot in black squares in Fig.~\ref{fig:12}(a-b), together with the dynamics of the total A/B bath population in blue spheres/red triangles, respectively. We observe two different behaviours depending on the timescales:
\begin{itemize}
 \item For short times, the QE relaxes very slowly (as we already show in Fig.~\ref{fig:11}(c)) decaying mainly into the B sites, which is more clear in the zoom we make in Fig.~\ref{fig:12}(b). Moreover, inspecting the distribution of these modes in real space, shown in Fig.~\ref{fig:12}(c), we realize the decay into the B bath has a very anisotropic shape resembling the one of the BS of the FCC lattice at the upper edge. This is expected since the $\kk$ modes resonant at the QE frequency, $\Delta=0$, are the same ones that give rise to the divergence in the upper edge of the FCC lattice. One remarkable difference with respect to the FCC BS is that the spatial decay of the wavefunction, $|C_{B,\mathbf{n}}|$, along the main axis (see, e.g., Fig.~\ref{fig:12}(d)), does not show an exponential attenuation, but rather seems to follow a power law decay $\sim 1/n$, being $n$ the distance from the QE. This behaviour at short times resembles the one of 2D Dirac cone, where a quasiBS emerges in the B/A lattice for a QE 
coupled to the A/B lattice site with a power-law localization of its wavefunction.

\item For long-times, on the contrary, the dynamics differs significantly from the 2D situation, where the QE decay freeze around a constant value~\cite{gonzaleztudela18c}. Here, instead, the excitation from the QE gets completely transferred to the A bath after a certain time. This can be understood from the existence of a collective mode of the $A_0$-bath at zero energy, which is able to resonantly transfer excitation from the QE to the bath. This mode can be defined as:
\begin{equation}
 A_0^\dagger =\frac{1}{\sqrt{N_{A_0}}}\sum_\kk a^\dagger_\kk\,.
\end{equation}
where $N_{A_0}$ is the number of $\kk$-modes which satisfy $|f(\kk)|^2=0$, which can be shown be given by the same closed contour composed of $12$ lines that we explained to describe the upper BS of the FCC lattice. Thus, this results into a resonant interaction between the QE and the collective mode of the type: $g_A(\sigma_{eg}A_0+\mathrm{h.c.})$, which gives rise to Rabi oscillations between the QE and $A_0$, with frequency:
\begin{equation}
 g_A\approx \frac{\sqrt{3}g}{N}
\end{equation}
which reproduces the behaviour of Fig.~\ref{fig:12}(a). The spatial distribution of this mode is plotted in Fig.~\ref{fig:12}(e), for a time $TJ=9N$, where most of the population has been transferred to the $A$ bath. Since the $g_A\propto 1/N$, in the thermodynamic limit this contribution vanishes and eventually only the relaxation from the B bath appears.

\end{itemize}

To conclude this section, let us remind that all the results with finite baths are obtained using periodic boundary conditions. Although some of the results may vary with different boundary conditions, e.g., the emergence of the zero energy mode, for the sake of concreteness we leave the detailed discussions with other boundary conditions for further works.

\subsection{Many QEs}

\begin{figure}
\centering
\includegraphics[width=0.4\textwidth]{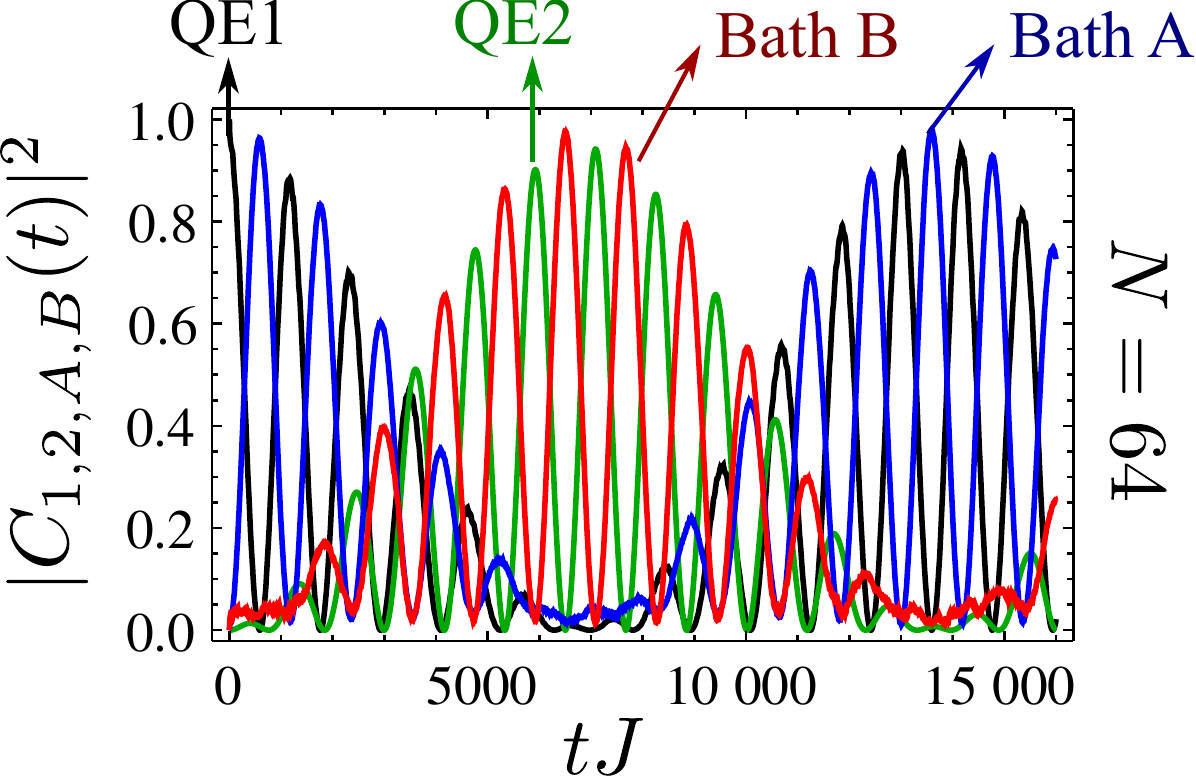}
\caption{$|C_{1,2}(t)|^2$ [$|C_{A,B}(t)|^2$] in solid black/green [red/green] lines for two QEs coupled with $g=0.1J$ and $\Delta=0$ to the A/B lattices of diamond with relative position $\nn_{AB}=(1,0,0)$, and bath size $N=64$.}
\label{fig:13}
\end{figure}

As a final illustration, let us consider how the non-trivial dynamics emerging at $\Delta=0$ translates to the situation where many QEs are interacting with the diamond bath. In particular, we consider a situation where two QEs, 1 and 2, are coupled respectively to the A/B lattice. Then, we set the first QE in the initial state, and study whether the excitation gets coherently transferred to the second one. In Fig.~\ref{fig:13} we show the dynamics of the two QEs, $|C_{1,2}(t)|^2$ in solid black/green, and of the total bath population $|C_{A,B}(t)|^2$ in the A/B lattices in blue/red, respectively. In analogy to what happened in the single QE situation, the behaviour occurs in two different timescales: the initially excited QE starts oscillating back and forth with the collective bath mode it is coupled to. However, as time passes there is a small fraction of population being transferred to the second QE (and its corresponding bath), until it gets transferred completely.

From here, there are many research directions to continue exploring, such as what happens in the limit $N\rightarrow\infty$, when the coupling to the collective zero-energy mode vanishes, or when QEs couple to the same sublattice. We leave them open for further works.

\section{Bath implementation with optical lattices  \label{sec:OLimplementation}}

In Section~\ref{sec:OL} we introduced the general ideas on how to simulate quantum optical phenomena with cold atoms in state-dependent optical lattices based on Ref.~\cite{devega08a}. In the original proposal, however, the bath was considered to be free-particles with energy dispersion $\omega(\kk)\propto |\kk|^2$. To observe the phenomenology explored in Sections~\ref{sec:CS}-\ref{sec:diamond}, the key ingredient is to be able to engineer more complicated bath energy dispersions, $\omega(\kk)$, like the ones considered along the manuscript.  Some of the bath geometries, like CS lattices, are straightforward to generate using three retro-reflected laser beams and they are used nowadays in most of the 3D state-of-art experiments~\cite{bloch08a}. The other geometries, however, have received much less attention~\cite{petsas94a,boretz15a,lang17a} such that it is still worth revisiting them and characterize their band structure.

In this Section: i) we provide the laser configurations to obtain the lattices explored along this manuscript; ii) we calculate the associated band structure of the laser configurations considered; iii) we analyze to which extent the dynamics predicted in this manuscript could be observed in these setups by studying the associated energy scales and potential problems, such as the emergence of longer-range hopping, which may alter the density of states of the ideal models. We focus on the study of the bath density of states since its non-analytical behaviour is crutial in most of the phenomenology predicted in the manuscript.  Here, we only characterize the bath energy dispersion and its hopping rates, since the atomic state playing the QE role is supposed to be in a deep lattice trap where all tunnelings are very much suppressed.

We also acknowledge there will be other experimental imperfections that may affect the predicted dynamics, e.g., decoherence. These effects have been studied for lower dimensional baths, e.g., in Refs.~\cite{gonzaleztudela17b,gonzaleztudela18c}, where the rule of thumb to observe the phenomena was that it occurs in timescales faster than the decoherence ones, something that is within the reach with these cold atoms setups. Finite size effects or the presence of defects may also renormalize the non-analytical features of the density of states, as we will show along this section and it has already been considered in the literature (see, e.g., Refs~\cite{antezza09a,antezza13a}). These imperfections generate as well a competing timescale with the observation of the desired features. However, we have already seen in this manuscript with exact simulation with finite lattices that the observation of the non-trivial dynamics is possible even with moderate system sizes ($N=256$) by choosing the QE-bath coupling 
appropriately.  Since the goal of the manuscript is to uncover new phenomena, we leave for future works the systematic characterization of these and other experimental restrictions.

\subsection{Tools \& Analysis \label{subsec:tools}}

\emph{Atomic optical potentials.}
The interaction of one or many lasers, with a given wavelength $\lambda$, with an atomic optical transition generates optical potentials which can be generally written as~\cite{grimm00a}:
\begin{align}
 \label{eq:oL}
 V(\RR)=V_0 I(\RR)\,,
\end{align}
where $V_0$ is the overall amplitude of the optical potential, and $I(\RR)$ its intensity profile. The parameter $V_0$ can be controlled in magnitude and sign through both the laser intensities and detunings with respect to the optical transitions. Along this manuscript, we will assume $V_0$ to have units of energy, such that $I(\RR)$ is dimensionless. The intensity profile, $I(\RR)$, depends on both the atomic polarizability tensor and its interplay with the total electric field. For simplicity, we assume to work in the regime where the polarizability tensor is isotropic and the intensity profile is just given by the total electric field profile, $|\EE(\RR)|^2$ resulting from the interference of the different lasers interacting with the atom. 

The first task in the following Sections will be to find laser configurations that give rise to optical potentials with the same Bravais structure than the bath geometries depicted in Figs.~\ref{fig:3}(a-d). For that purposes, we use two experimental resources:

i) On the one hand, we exploit the interference between several laser fields with the same wavelength (or equivalently frequency). Each laser field is characterized by its amplitude/polarization/propagation vector $\{E_i,\hat{\varepsilon}_i,\pp_i\}$, that is, they have associated a vector electric field : $\EE_i(\rr)=e^{i\pp_i\cdot\rr}E_i\hat{\varepsilon}_i$. Since we assume they all have the same frequency, the interference leads to an electric field intensity which reads:
\begin{align}
  |\EE(\RR)|^2=\sum_{i}|E_i|^2+2\sum_{i>j}E_i E_j\Big[\mathrm{Re}(\hat{\varepsilon}_i\hat{\varepsilon}^*_j)\cos\left((\pp_i-\pp_j)\cdot \RR)\right) -\mathrm{Im}(\hat{\varepsilon}_i\hat{\varepsilon}^*_j)\sin\left((\pp_i-\pp_j)\cdot \RR)\right)\Big]\,.\label{eq:intsame}
\end{align}

It contains both a constant shift of the potential coming from the self-interference of each electric field, and a position dependent contribution coming from the interference of each pair of electric fields. 

ii) Since sometimes these patterns will not be enough, we will need to add up the contribution of other laser fields with different wavelength (and frequencies, $\omega_\alpha$) such that they result in a time-independent averaged potential:
\begin{equation}
\label{eq:intE}
 |\EE(\RR)|^2\approx \sum_{\alpha}|\EE_\alpha(\RR)|^2\,,
\end{equation}
where $E_\alpha(\RR)$ is the total electric coming from the interference of the laser fields with the same wavelength $\omega_\alpha$ as given by Eq.~\ref{eq:intsame}. The cross-interference between the lasers with different frequencies averages out in our timescales of interest when $\Delta\omega_{\alpha\beta}=\omega_\alpha-\omega_\beta$ is bigger than any parameter of our simulated Hamiltonian ($\lesssim 10$ kHZ), which is the regime we will assume to be working in. Since:
\begin{equation}
 \Delta\omega_{\alpha\beta}=2\pi\times c \frac{\Delta \lambda}{\lambda_\alpha \lambda_\beta}\,,
\end{equation}
for $\lambda_i\sim 500$ nm, we have that $\Delta  \omega\approx 2\pi\times 10^9 \Delta\lambda$ KHz/nm, such that they can be easily satisfied even for very closely spaced wavelengths, which we consider to the same for the purposes of the defining the lattice geometry.

\emph{Characterizing bath structure.}
Once we find an appropriate laser configuration for each bath, we will characterize the emergent atomic dynamics for atoms hopping in a given potential $V(\RR)$. This is important since having the same periodicity as the original model does not guarantee that the dynamics will be described by the same energy dispersion. In particular, longer range hoppings may emerge that deviate the dynamics from the nearest neighbours descriptions that we considered in the previous Sections.

The Hamiltonian describing the dynamics of atoms hopping in these optical lattices can be generally written~\cite{bloch08a}:
\begin{align}
 \label{eq:hamOL}
 \hat{H}=\left[-\frac{\hbar^2}{2 M} \nabla^2+V(\mathbf{R}) \right]\,,
\end{align}
where $M$ is the atomic mass and $\nabla^2=\frac{\partial^2}{\partial x^2}+\frac{\partial^2}{\partial y^2}+\frac{\partial^2}{\partial z^2}$ is the Laplacian in cartesian coordinates. To make the problem adimensional, it is convenient to use $\lambdabar=\lambda/(2\pi)$ and the recoil energy $E_R=\frac{\hbar^2}{2M \lambdabar^2}$ as the natural units of length and energy respectively~\footnote{The recoil energy may depend on the particular laser configuration considered, but we will keep the same notation for it.}. With these units, the Hamiltonian of Eq.~\ref{eq:hamOL} is reexpressed:
\begin{align}
 \label{eq:hamOL2}
 \hat{H}=\left[-\nabla^2+V(\RR) \right]\,,
\end{align}
where for simplicity we have kept the same notation for $\hat{H},V(\RR)$ and $V_0$, even though they are all now written in units of the recoil energy. Since we are interested in calculating the lowest energy part of the spectrum, we assume to have translational invariant system to apply Bloch Theorem~\cite{wannier37a}. Under this assumption, $\kk$, is a good quantum number such that the eigenfunctions can always be written as: $\ket{\Psi(\RR)}=e^{i \kk \cdot \RR} \ket{u_{\kk}(\RR)}$, where $\ket{u_\kk(\RR)}$ are the so-called Bloch-modes. Using this ansatz, the eigenvalue equation transforms to:
\begin{align}
 \label{eq:hamOL2}
 \hat{H}_\kk\ket{u_{\kk,n}(\RR)}=E_{\kk,n}\ket{u_{\kk,n}(\RR)}\,,
\end{align}
where the index $n$ denotes the different bands, $E_{\kk,n}$, appearing in the model. The $\kk=\sum_{j=1}^3 k_j \dd_j$, where $\dd_j$ are the reciprocal primitive vectors of each lattice used to define the periodic boundary conditions of the problem. Since the reciprocal primitive vectors are different for each lattice considered, the shape of $\hat{H}_\kk$ varies for every lattice geometry, like we will show explicitly in the next Sections. Finally, since both the potential $V(\RR)$ and the Bloch mode, $\ket{u_{\kk,n}(\RR)}$, have the same spatial periodicity, it is convenient to expand them in terms of a finite number of reciprocal wavevectors:
\begin{align}
 \label{eq:exp}
 V(\RR)&=\sum_{\qq} V_\qq e^{i \qq \cdot \RR}\,.\\
 \ket{u_{\kk,n}(\RR)}&=\sum_{\qq}^{|q_j|<q_\mathrm{max}} C_{\kk,n,\qq} e^{i \qq \cdot \RR}
\end{align}
where $q_\mathrm{max}$ is the numerical cut-off to expand the Bloch modes in the reciprocal space. Thus, to obtain $E_{\kk,n}$ for each $\kk$-point, one needs to solve a eigenvalue equation for matrices of size $(2q_\mathrm{max}+1)^3$. This problem simplifies a lot for the case of separable potentials, i.e., $V(\RR)=V(x)+V(y)+V(z)$, where one can solve directly the 1D problem where the size of the the matrix is just $(2q_\mathrm{max}+1)$.  Except for the diamond lattice case, that will be treated separately in Section~\ref{subsec:diam}, we focus only on the lowest energy band such that from now on we drop the index $n$ of the discussion.  

Like we mentioned in the beginning of this Section, we will calculate the following figures of merit to estimate to which extent the physics predicted in this manuscript can be observed within optical lattices:
\begin{itemize}
 \item On the one hand, we calculate the strength of the atom (nearest neighbour) hoppings. It can be easily shown that the hopping rate between two atoms separated a distance $\nn$ can be obtained from by Fourier transforming $E_{\kk}$:
 \begin{align}
  \label{eq:hop}
  J_\nn=\frac{1}{N^3}\sum_{\kk} E_{\kk} e^{-i\kk\cdot\nn}
 \end{align}
 where $N$ is the (linear) number of sites that the use to discretize the $k_i$ variables.  
 
 \item On the other hand, we also calculate the numerical density of states of the bath, by discretizing the energy space, which can be calculated from $E_{\kk}$, which can be calculated as follows:
 \begin{equation}
  \label{eq:dosnum}
  D(\omega_n)=\frac{1}{N^3}\sum_{\kk} \Theta(\omega_{n-1} <E_\kk <\omega_n)
 \end{equation}
where $\Theta(x)$ is the Heaviside function, $\omega_n=\mathrm{min}E_\kk+n\left( \frac{\mathrm{max}E_\kk-\mathrm{min}E_\kk}{N_\omega}\right)$, being $N_\omega$ the number of steps in which we discretize the frequency space and $n=1,\dots N_\omega$. We will be especially interested in checking whether the non-analytical features observed in Figs.~\ref{fig:3}(e-h) survive in the real lattices.
\end{itemize}

After we have given the tools and prescriptions to analyze the problems, let us study the results obtained for the different lattice geometries considered.

\subsection{Cubic Simple lattices \label{subsec:CS}}

This lattice is simple to generate~\cite{bloch08a}: one just needs three retro-reflected lasers, with orthogonal polarizations, propagating in the x,y,z directions, that is, $E_i=E_0$, $\pp_{1,2/3,4/5,6}=\pm \frac{2\pi}{\lambda}\hat{\bf{e}}_{x/y/z}$ and $\hat{\varepsilon}_{1,2/3,4/5,6}=\hat{\bf{e}}_{z/x/y}$. The interference of these six electric fields results in a potential:
\begin{align}
 \label{eq:potCS}
 V_\CS(\RR)=V_0\left[\cos^2(x)+\cos^2(y)+\cos^2(z)\right]\,.
\end{align}

Since the potential and the kinetic energy terms are both separable in this case, one can treat the problem of each dimension separately. This allows us to make calculations for very large lattices with little numerical effort.

\begin{figure}
\centering
\includegraphics[width=0.666\textwidth]{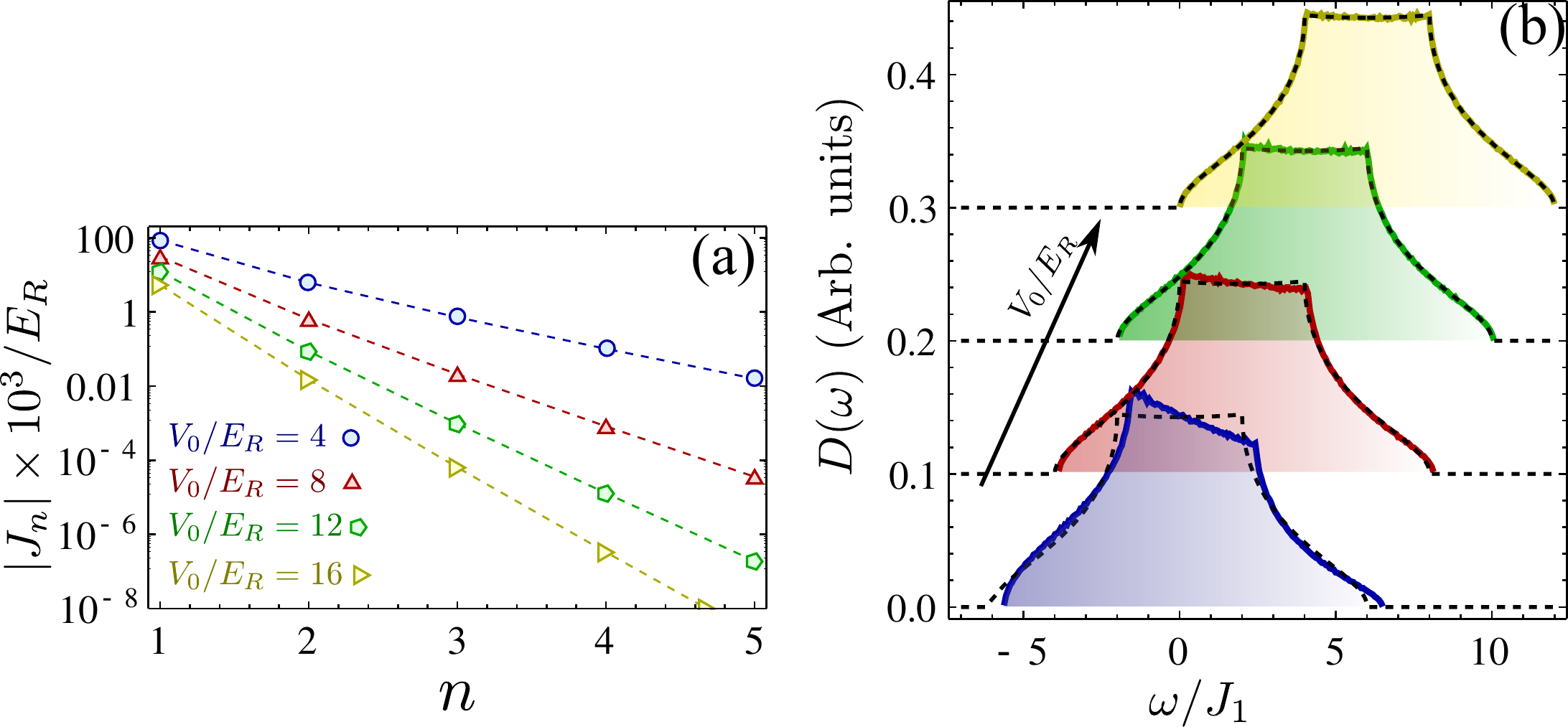}
\caption{CS bath: (a) Hopping rate to the $n$-th neighbour, $|J_n/E_R|$, for several $V_0/E_R=4$ (blue spheres), $8$ (red triangles), $12$ (green pentagons), $16$ (yellow triangles).  (c) Bath density of states for the different $V_0/E_R$ discussed in the panel (a). In dashed black lines we plot the expected result for the nearest neighbour model in the thermodynamic limit. Numerical simulations done for a bath of linear size $200$ and momentum cut-off $q_\mathrm{max}=10$.}
\label{fig:14}
\end{figure}

The first thing we are interested in is the scaling of the $n$-th neighbour hopping rate, that we denote as $J_n$, with the lattice potential $V_0$. In this geometry there are six nearest neighbours at positions $(\pm 1,0,0),(0,\pm 1,0),(0,0,\pm 1)$. Due to the separability of the potential, the longer range hoppings also appear only in these directions, such that we can denote as $J_n$ to the hopping rate of the $n$-th neighbour at position, e.g., $(n,0,0)$. In Fig.~\ref{fig:14}(a) we plot the result of applying Eq.~\ref{eq:hop} with these positions for a system with $N=200$ lattice sites and four different values of $|V_0/E_R|=4,8,12,16$. As expected, the larger $V_0$, the smaller $J_1$ since the Wannier functions in each site get more localized and their overlap decrease. The longer range hoppings decay quickly with the distance due to the exponential localization of the Wannier wavefunctions.

Finally, in Fig.~\ref{fig:14}(b) we plot in solid lines the numerical density of states for the lattice depths considered in the panels (a-b), together with the expected one from the nearest neighbour model (in dashed black lines). For the shallowest lattice considered, in blue, we observe how an asymmetry between the two kinks in the middle of the band appear. However, as the potential depth increases one quickly recovers the nearest-neighbour density of states. Remarkably, the non-analyticities giving rise to non-Markovian phenomena seem to survive even in the cases where the model deviates from the ideal nearest neighbour description.

\subsection{Body-Centered Cubic lattices \label{subsec:BCC}}

The minimal configuration to engineer a BCC optical potential requires four lasers in an umbrella-like or XY-YZ configuration, as discussed in Ref.~\cite{petsas94a}. Instead, we prefer to choose a variation of the proposal of Ref.~\cite{boretz15a} which uses more laser fields, since it can be implemented through retro-reflection. The proposal of Ref.~\cite{boretz15a} consists in using three pairs of laser fields with the same amplitude $E_i=E_0$ propagating in three orthogonal directions, $\pp_{1,2/3,4/5,6}=\pm \frac{2\pi}{\lambda}\hat{\bf{e}}_{x/y/z}$, but with three non-orthogonal polarizations, $\hat\varepsilon_{1,2}=(\hat{\mathbf{e}}_y+\hat{\mathbf{e}}_z)/\sqrt{2}$,$\hat\varepsilon_{3,4}=(\hat{\mathbf{e}}_x+\hat{\mathbf{e}}_z)/\sqrt{2}$ and $\hat\varepsilon_{5,6}=(\hat{\mathbf{e}}_x+\hat{\mathbf{e}}_y)/\sqrt{2}$. The resulting potential from the interference of these fields is given 
by:
\begin{align}
 \label{eq:VoptBCC}
 V_\BCC(\RR)&=V_\CS(\RR)+V_0\Big[\cos(x)\cos(y)+\cos(y)\cos(z)+\cos(x)\cos(z)\Big]\,.
\end{align}

\begin{figure}
\centering
\includegraphics[width=1.0\textwidth]{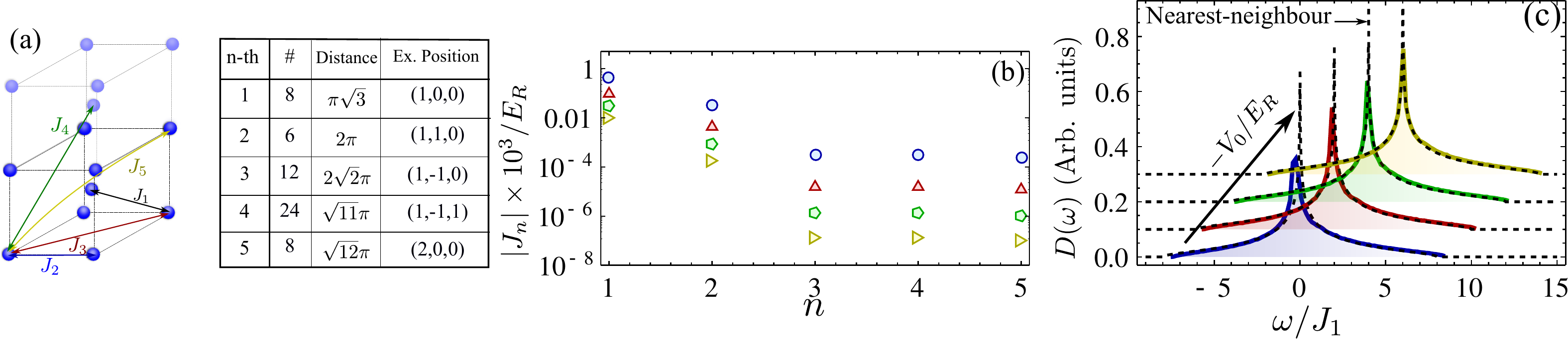}
\caption{BCC bath: (a) Sketch and summary of longer range neighbours. Table: number of $n$-th neighbours, distance and an example position of one of the $n$-th neighbour. (b) $n$-th neighbour atom hopping, $|J_n/E_R|$ for several $V_0/E_R=-3$ (blue spheres), $-4$ (red triangles), $-5$ (green pentagons), $-6$ (yellow triangles).  (c) Bath density of states for the different $|V_0/E_R|$ discussed in the panel (b). In dashed black lines, we plot the expected result for the nearest neighbour model in the thermodynamic limit.  Numerical simulations done for a bath of linear size $2^5$ and momentum cut-off $q_\mathrm{max}=7$.}
\label{fig:15}
\end{figure}

The primitive real/reciprocal vectors of this potential are: $\cc_{1}=\pi(1,1,-1)$, $\cc_{2}=\pi(1,-1,1)$ and $\cc_{3}=\pi(-1,1,1)$/$\dd_{1}=\frac{1}{2\pi}(1,1,0)$, $\dd_{2}=\frac{1}{2\pi}(1,0,1)$ and $\dd_{3}=\frac{1}{2\pi}(0,1,1)$, respectively, which can be shown to expand a BCC lattice. The summary of the number, distance and position of nearest neighbours is shown in Fig.~\ref{fig:15}(a). We use $J_n$ to denote the hopping rate of the $n$-th neighbour.

To calculate the band structure of this potential it is convenient to express the position/momenta in the eigenvalue equation in terms of the primitive vectors of real/reciprocal space. For example, the Laplacian, $\nabla^2$, written in terms of $\nn=\sum_j n_j \cc_j$ is given by:
\begin{equation}
\label{eq:nablabcc}
\nabla^2=\frac{2}{(2\pi)^2}\Big[\frac{\partial^2}{\partial n_1^2}+\frac{\partial^2}{\partial n_2^2}+\frac{\partial^2}{\partial n_3^2}+\frac{\partial^2}{\partial n_1\partial n_2}+\frac{\partial^2}{\partial n_2\partial n_3}+\frac{\partial^2}{\partial n_1\partial n_3}\Big]\,.
\end{equation}

The potential in primitive coordinates is simply obtained by:
\begin{equation}
 V_\BCC(x,y,z)=V_\BCC\left(\pi(n_1+n_2-n_3),\pi(n_1-n_2+n_3),\pi(-n_1+n_2+n_3)\right)\,.
\end{equation}

With this change of variables, one can calculate $E_\kk$ using the prescriptions explained in Section~\ref{subsec:tools}. Since the potential is not separable and the calculations are more demanding we calculate the band structure for smaller system sizes than the CS lattice, and then extrapolate the results to larger lattices to obtain a smooth density of states. 

We numerically calculated $E_\kk$ using the potential $V_\BCC(\RR)$ and realize that it deviates from the energy the energy dispersion of the nearest neighbour model (not shown). The underlying reason is that the potential in the direction of next-nearest neighbour is shallower than in the nearest neighbour direction, such that it still has a big effect in spite of the larger distance. To solve this problem we propose to use another set of 3 retro-reflected lasers with different frequency but virtually indistinguishable wavelength, as explained in Section~\ref{subsec:tools},  and with negative $V_0$, such that they cancel the $V_\CS(\RR)$ contribution of $V_\BCC(\RR)$. In that case the potential finally reads:
\begin{align}
 \label{eq:VoptBCC2}
 V_{\BCC,2}(\RR)=V_0\Big[\cos(x)\cos(y)+\cos(y)\cos(z)+\cos(x)\cos(z)\Big]\,.
\end{align}

Another possibility to obtain $V_{\BCC,2}(\RR)$ consists in summing three different  sets of lasers, with the same amplitude $E_0$ and slightly different frequencies, with the following propagation/polarization vectors: The first set is composed by $\pp_{1/3}=\frac{\pi}{\lambda}(1,\pm 1,0)=-\pp_{2/4}$, with polarizations $\hat\varepsilon_{1,2/3,4}=\frac{1}{\sqrt{2}}(1,\mp 1,0)$. The second set is composed by $\pp_{5/7}=\frac{\pi}{\lambda}(1,0,\pm 1)=-\pp_{6/8}$, with polarizations $\hat\varepsilon_{5,6/7,8}=\frac{1}{2}(1,0,\mp 1)$. The third set is finally composed by $\pp_{9/11}=\frac{\pi}{\lambda}(0,1,\pm 1)=-\pp_{10/12}$, with polarizations $\hat\varepsilon_{9,10/11,12}=\frac{1}{\sqrt{2}}(0,1,\mp 1)$. The time-averaged potential resulting from the sum of the interference of each set leads to:
\begin{align}
 \sum_\alpha |\EE_\alpha(\RR)|^2&\propto \cos(x+y)+\cos(x-y)+\cos(x+z) +\cos(x-z)+\cos(y+z)+\cos(y-z)\nonumber \\
 & = \cos(x)\cos(y)+\cos(x)\cos(z)+\cos(z)\cos(y)\,.
\end{align}

Irrespective of the method used to obtain $V_{\BCC,2}(\RR)$, we calculate the lowest energy band of the atoms hopping in this potential and summarize the results in Figs.~\ref{fig:15}(b-c). In Fig.~\ref{fig:15}(b) we plot the scaling of the $n$-th nearest neighbour hopping for several  lattice depths $V_0$ given in the legend, where we observe that indeed $|J_1|\gg |J_{n>1}|$ for large lattice depths.  Finally, in Fig.~\ref{fig:15}(c) we plot the numerical density of states for the lattice depths considered in the previous panels. For shallow lattices, the longer range hoppings renormalize the divergence of the density of states in the middle of the band. For the deeper lattices, the results converge to the ones expected from the nearest neighbour description.

\subsection{Face-Centered Cubic lattices \label{subsec:FCC}}

There are several ways of obtaining FCC optical potentials~\cite{petsas94a,lang17a}. In Ref.~\cite{petsas94a} it was proposed to use a minimal configuration using 4 lasers, whereas in Ref.~\cite{lang17a} they use 3 retro-reflected beam to obtain a FCC geometry which, unfortunately, could not be symmetric in X/Y/Z. We propose instead to use 4 retro-reflected laser beams with the same amplitude and slightly different frequency (but similar wavelength) as explained in Section~\ref{subsec:tools}. The propagation/polarization vectors of the laser field (and their reflection) are: $\pp_{1/2}=\pm \frac{\pi}{\lambda}(1,1,1)$, $\pp_{3/4}=\pm \frac{\pi}{\lambda}(1,1,-1)$, $\pp_{5/6}=\pm \frac{\pi}{\lambda}(1,-1,1)$, $\pp_{7/8}=\pm \frac{\pi}{\lambda}(-1,1,1)$ and  $\hat\varepsilon_{1,2}\propto (1,-2,1)$, $\hat\varepsilon_{3,4}\propto (1,1,2)$
, $\hat\varepsilon_{5,6}\propto (1,2,1)$,  $\hat\varepsilon_{7,8}\propto (2,1,1)$. The resulting potential, up to a constant shift, reads:
\begin{align}
\label{eq:OptFCC}
V_\FCC(\RR)=V_0 \cos(x)\cos(y)\cos(z)\,.
\end{align}

\begin{figure}
\centering
\includegraphics[width=1.0\textwidth]{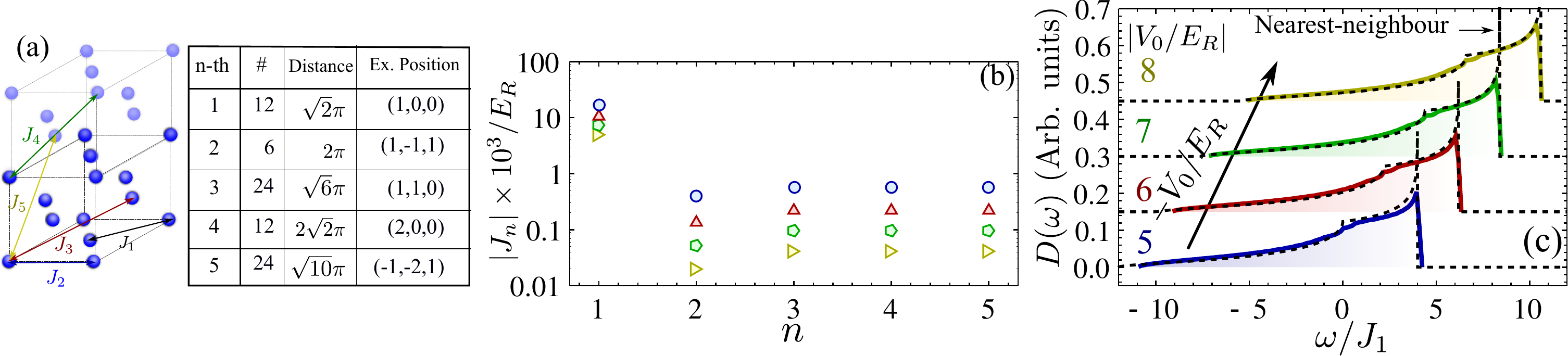}
\caption{FCC bath: (a) Sketch and summary of longer range neighbours. Table: number of $n$-th neighbours, distance and an example position of one of the $n$-th neighbour. (b) $n$-th neighbour atom hopping, $|J_n/E_R|$ for several $V_0/E_R=-5$ (blue spheres), $-6$ (red triangles), $-7$ (green pentagons), $-8$ (yellow triangles).  (c) Bath density of states for the different $|V_0/E_R|$ discussed in the panel (b). In dashed black lines, we plot the expected result for the nearest neighbour model in the thermodynamic limit.  Numerical simulations done for a bath of linear size $2^5$ and momentum cut-off $q_\mathrm{max}=7$.}
\label{fig:16}
\end{figure}

The primitive real/reciprocal vectors of this potential are: $\cc_{1}=\pi(1,1,0)$, $\cc_{2}=\pi(1,0,1)$ and $\cc_{3}=\pi(0,1,1)$/$\dd_{1}=\frac{1}{2\pi}(1,1,-1)$, $\dd_{2}=\frac{1}{2\pi}(1,-1,1)$ and $\dd_{3}=\frac{1}{2\pi}(-1,1,1)$, respectively. As summarized in Fig.~\ref{fig:16}(a), this lattice geometry is characterized by having 12 nearest neighbours. The number, distance, and position of the longer-range neighbours is also given in the table of Fig.~\ref{fig:16}(a) up to the fifth neighbour.

 Like we did for the BCC potential, to calculate the band structure it is convenient to express the position/momenta in the eigenvalue equation in terms of the primitive vectors. For example, the Laplacian, $\nabla^2$, in terms of $\nn=\sum_j n_j \cc_j$ is given by:
\begin{align}
\label{eq:nablabcc}
\nabla^2=\frac{2}{(2\pi)^2}\Big[&\frac{3}{2}\frac{\partial^2}{\partial n_1^2}+\frac{3}{2}\frac{\partial^2}{\partial n_2^2}+\frac{3}{2}\frac{\partial^2}{\partial n_3^2}-\frac{\partial^2}{\partial n_1\partial n_2}-\frac{\partial^2}{\partial n_2\partial n_3}-\frac{\partial^2}{\partial n_1\partial n_3}\Big]\,.
\end{align}

The potential written in primitive coordinates can be obtained replacing: $x=\pi(n_1+n_2), y=\pi(n_1+n_3), z=\pi(n_2+n_3)$. With this change of variables, we calculate $E_\kk$ using the prescriptions explained in Section~\ref{subsec:tools}.  The results are summarized in Fig.~\ref{fig:16}, where we plot both the scaling of the $n$-th neighbour hopping for several $V_0/E_R$ in Fig.~\ref{fig:16}(b), and the extrapolated numerical density of states in Fig.~\ref{fig:16}(c). We observe that the numerical density of states approaches the nearest neighbour one like we desired.

\subsection{Diamond lattices \label{subsec:diam}}

We already explained in Section~\ref{sec:bath} that the diamond lattice is formed by two interspersed FCC lattices displaced by a vector $\mathbf{f}=\left(\frac{1}{4},\frac{1}{4},\frac{1}{4}\right)$. Using this intuition, one could build an optical potential by doubling the configuration of the FCC lattice to make a second displaced potential using the trick of making two slightly detuned sets of lasers. The final optical potential will be given by the sum of the two displaced potentials:
\begin{align}
 V_\mathrm{diam}(\RR)=V_\FCC(\RR)+V_\FCC(\RR+\mathbf{f})\,,
 \end{align}

\begin{figure}
\centering
\includegraphics[width=0.999\textwidth]{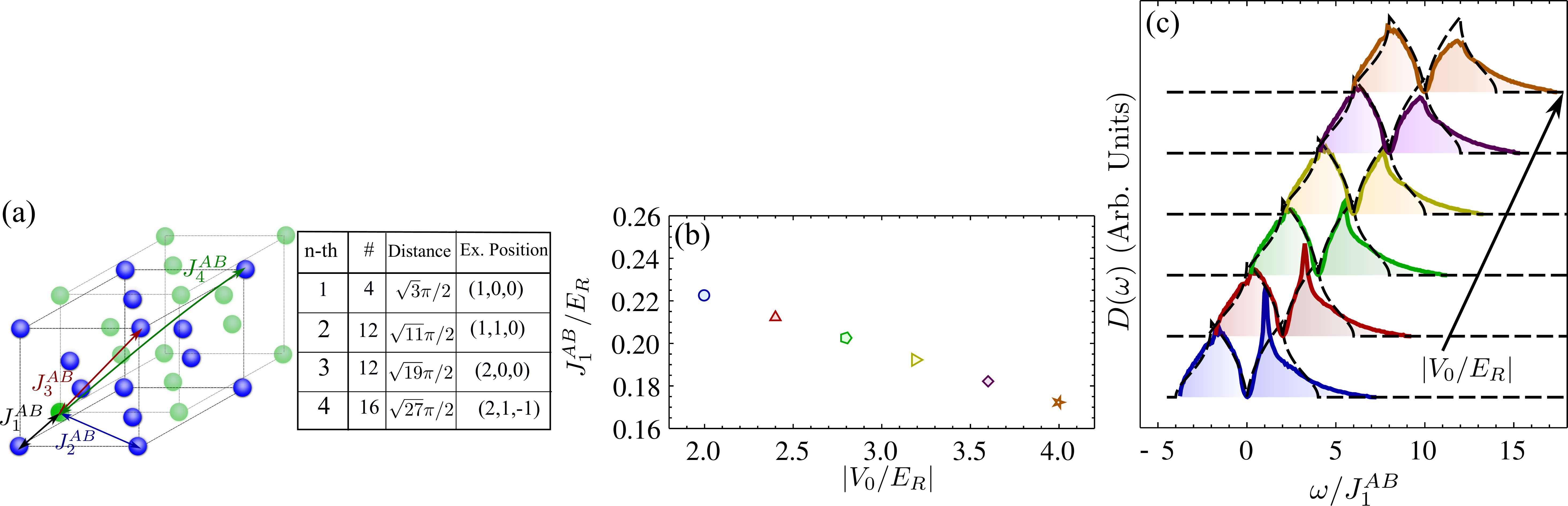}
\caption{Diamond bath: (a) Sketch and summary of longer range neighbours between the AB lattices. Table: number of $n$-th neighbours, distance and an example position of one of the $n$-th neighbour.  (b) Scaling of the dominant nearest neighbour contribution, $J^{AB}_1$, for several lattice depth potentials $V_0/E_R$. (c) Bath density of states for the $V_0/E_R$ considered in panel (b). In dashed black lines we plot the expected result for the nearest neighbour model in the thermodynamic limit.  Numerical simulations done for a bath of linear size $2^5$ and momentum cut-off $q_\mathrm{max}=7$. }
\label{fig:17}
\end{figure}

The primitive vectors for the real/reciprocal space are those of the FCC potential, as well as the nearest neighbours within the same sublattice. Besides them, there are extra neighbours emerging from the coupling to the other sublattice sites, depicted and summarized in Fig.~\ref{fig:17}(a). 

Furthermore, another difference of this situation with respect to the case of a simple Bravais lattice is that we need to calculate not only the lowest, but also the first excited band, that we denote as $E_{1,\kk}$ and $E_{2,\kk}$ to characterize the behaviour of the system. In this type of models with a superlattice the Wannier function are not unambiguously defined~\cite{marzari97a}, however, the maximally localized ones lead to the following formulas for the tunneling within the AA (BB) or AB lattice sites~\cite{uehlinger14a}:
\begin{equation}
\label{eq:tunAA}
J^{\mathrm{AA/AB}}_\nn=\frac{1}{2 N^3}\sum_\kk e^{-i\kk \cdot \nn}\left(E_{1,\kk}\pm E_{2,\kk}\right)\,.
\end{equation}

For the sake of concreteness, we will only plot here the scaling of the most relevant nearest neighbour coupling, $J^{\mathrm{AB}}_1$, for several $V_0/E_R$ in Fig.~\ref{fig:17}(b), and the corresponding density of states in Fig.~\ref{fig:17}(c), calculated extrapolating the results of a small system size simulation like we did in the previous cases. Differently from what happened in the other types of baths, with the potential configuration we propose there is no range of $V_0/E_R$ that approximates $D_\mathrm{diam}(\omega)$ in all spectral regions. While the lowest band is reasonably well approximated for all the lattice depth range considered, the upper one deviates significantly for the idealized nearest neighbour model, being more similar around $V_0/E_R\sim 3$. However, it is worth emphasizing that the regions around $\omega\approx 0$ does indeed show a singular bandgap, which points to the possibility of observing the most distinctive dynamics of this type of reservoirs.

\section{Conclusions \& Outlook \label{sec:conclusion}}

Summing up, we have systematically characterized the quantum dynamics of QEs coupled to several 3D structured reservoirs with different qualitative features in their density of states. Through exact calculations, we predict the emergence of several phenomena beyond the traditionally considered band-edge related effects, such as:
\begin{itemize}
 \item Long-lived reversible dynamics for single QEs spectrally tuned within band frequencies in CS lattices.
\item Directional emission and the emergence of perfect subradiant states in BCC lattices.
\item Robust 3D anisotropic bound states which survive irrespective of the QE spectral detuning/coupling for FCC lattice. This is in stark contrast to standard 3D bound states appearing in isotropic band-edges, which for a fixed $g/J$ merge into the scattering spectrum for a critical $\Delta$.

\item Through these robust bound states the FCC bath also mediates QE dipole-dipole interactions only in certain directions. The spatial decay of the interactions in these directions can be numerically fitted to: $\sim e^{-d\sqrt{\delta}n }/\sqrt{\delta n}$, very different from the isotropic Yukawa type interactions of standard 3D reservoirs.

\item Sub-exponential relaxation of single QEs tuned at the singular bandgap of diamond lattices. Furthermore, when many QEs are interacting with the bath at this frequency one observes reversible exchange of excitations with frequency scaling as $1/n$, being $n$ the distance between emitters.
\end{itemize}

Furthermore, we also propose a way how to observe these effects with cold atoms in state-dependent optical lattices based on the proposal of Ref.~\cite{devega08a}. More concretely, we i) provide the laser configurations to design optical potentials with the geometries of the bath considered, and ii) characterize their lowest energy band (or the two lowest energy bands for the case of diamond geometries). In particular, we study the associated timescales of the hopping models and to which extent their dynamics reproduce the physics of the nearest neighbour idealized models considered along the manuscript, taking their density of states as a figure of merit for the comparison. We conclude that the phenomena predicted along this manuscript is within the reach of this platform, that together with the recent experimental developments~\cite{krinner18a}, foreseeing the observation of non-trivial dynamics in near future experiments.

The work constitutes a solid basis for future investigation of 3D structured quantum optical systems and opens several research directions. From the fundamental point of view, natural extensions of this work are the study of the emergent phenomenology in the many excitation regime~\cite{shi18a}, exploring the interplay between driving and the anisotropic collective dissipation appearing in these systems which may lead to the emergence of many-body entangled steady states~\cite{ramos14a,pichler15a}, or the characterization of the phases emerging in the effective spin models. Another interesting direction consists in combining the ideas developed along this manuscript with the recent developments in confined photons in subwavelength atomic lattices~\cite{asenjogarcia17a,shahmoon17a,glaetzle17a,perczel17a}. From the more applied perspective, it will be interesting to find simpler laser configurations which give rise to the 3D models considered and make a more thorough characterization on the 
emergent dynamics with these potentials.

\section*{Acknowledgements}
The authors acknowledge the ERC Advanced Grant QENOCOBA under the EU Horizon 2020 program (grant agreement 742102). We also acknowledge enlightening discussions with J.~Arg\"uello-Luengo, E. S\'anchez-Burillo, S.~Blatt and I.~Bloch.

\appendix



\bibliographystyle{apsrev4-1}
\bibliography{Sci,books}

\end{document}